\documentclass[12pt,preprint]{aastex}
\pdfoutput=1

\begin{document}

\title{An Objective Survey of  Mpc-Scale Radio Emission in 0.03$<$z$<$0.3 Bright X-ray Clusters}
\shorttitle{Mpc-scale radio emission}
\author{Lawrence Rudnick\altaffilmark{1} and Jeffrey A. Lemmerman\altaffilmark{1}}
\altaffiltext{1}{University of Minnesota, 116 Church 
Street SE, Minneapolis, MN  55455;  corresponding author: larry@astro.umn.edu}

\begin{abstract}
We have performed the the first census of Mpc-scale radio emission to include control fields and quantifiable upper limits for bright X-ray clusters in the range 0.03$<$z$<$0.3 . Through reprocessing radio  images from the WENSS survey, we detect diffuse emission  from $\approx$30\% of the sample.   We find a correlation similar to  the well-studied relationship between radio halo and X-ray luminosities of the host cluster, but also find that large scale radio galaxy detections  follow a similar trend to that for radio halos.  With this quantitative study, we thus confirm the upper envelope to the radio luminosities for X-ray selected clusters, and the higher detection rates for diffuse radio emission (including halos, relics and radio galaxies) at X-ray luminosities above $\approx$10$^{45}$erg/s.   Beyond this, because many of the upper limits are only slightly below the detections, we can neither confirm nor refute the claims for a tight correlation between these radio halo and X-ray luminosities, i.e.  whether the halo luminosity function of X-ray clusters is bimodal (having  high/on and low/off states), or whether the radio luminosity can take on a wide range of values up to a maximum at each cluster X-ray luminosity.  The resolution of this issue, also recently addressed in a GMRT survey at 610~MHz, may provide a unique diagnostic for the timescales over which relativistic particles can be accelerated following cluster mergers. We discuss several important selection effects on radio vs. X-ray luminosity correlations, including surface brightness thresholds and non-X-ray-selected diffuse radio sources.  We also report several new detections of diffuse emission, including a Mpc-scale relic in RXJ1053.7+5450,  a possible halo/relic combination in Abell 2061, a serendipitous diffuse X-ray source associated with poor clusters in the Abell 781 field, and confirmation of very weak diffuse emission patches outside of Abell~2255. 
\end{abstract}


\section{ Introduction}
The first example of a Mpc-scale radio halo -- centrally located diffuse radio emission not associated with an individual AGN -- was detected in the Coma cluster by \cite{large}, although its nature was only described later by \cite{willson}.  Several halos were discovered shortly thereafter using the WSRT and single dish telescopes \citep{hanisch,noord}.  The first systematic search for halos in a large sample was done using the NRAO 300' telescope \citep{jaffe}, yielding upper limits for 610 MHz brightness of 0.8K - 7K, compared to the Coma brightness of 2.5K (also discovering, but not recognizing, the first ``relic" source, near Coma A).  Since that time, a number of other halo searches have been undertaken \citep{kemp01,giov99}, although upper limits are rarely reported.

Using only confirmed halo detections  \cite{liang} first showed a correlation between the radio and X-ray luminosities of galaxy clusters, arguing that the thermal electrons provided the seeds for the acceleration of relativistic particles.  Correlation studies and the physical models for halos  grew more sophisticated in the following years, as the apparent association with cluster mergers led to a possible energization source for Mpc halos \citep{burns95,ferr95, vent00, kemp01}.  Models have considered both  primary production of relativistic electrons through shocks and turbulence \citep[e.g.,][]{brun01} and secondary production from the longer-lived cosmic ray protons \citep{dennison, blasi99,mini01}.  Several different scaling relations have been studied, for example, radio luminosity with X-ray temperature and mass, and with halo size \citep[e.g.,][]{cass07}.  

However, the lack of reported upper limits for halo emission has remained.   \cite{ferr05} has noted that it is difficult to define upper limits for radio halos, because their size is not known {\it a priori}.  Without upper limits, it is not possible to establish whether there is a ``trigger''  (e.g. mergers) or equivalently, a threshold (e.g. mass) for halo formation. The alternative is that the energization of the relativistic plasma is a scale-free process, up to some maximum total available energy, and the non-detections form a continuous distribution up to the detection level.  After this paper was originally submitted, a series of analysis and data papers on a very sensitive survey for Mpc-scale emission in X-ray clusters was published by \cite{brun07, cass08, vent07,vent08} (hereafter, jointly GCLUS) for 0.2$<$z$<$0.4; at these  redshifts, 1~Mpc corresponds to several arcminutes, well within the angular scales accessible to the GMRT and other radio interferometers.

\begin{figure}[h]
\begin{center}
\includegraphics[width=12cm]{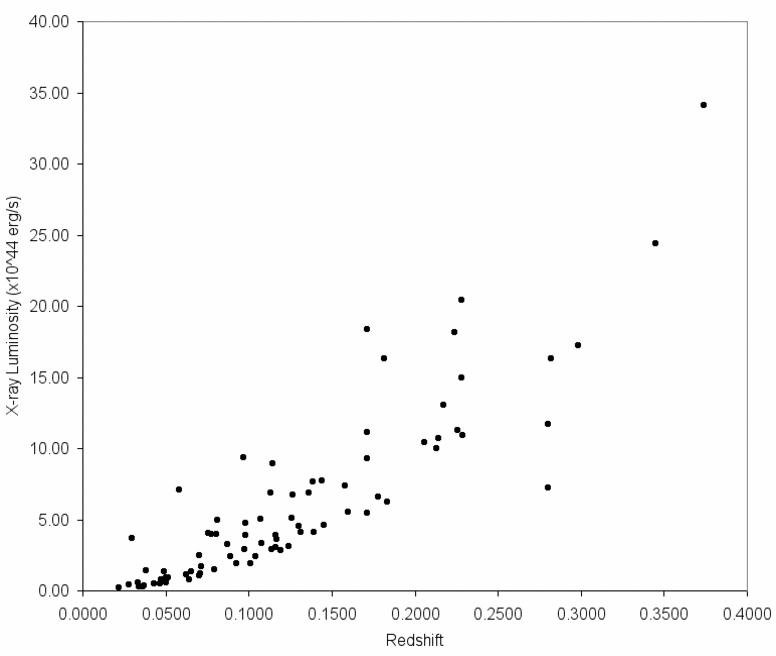}
\end{center}
\caption{X-ray luminosity as a function of redshift for the Ebeling et al. sample.  The point at z=0.28 below the rest of the distribution is from a second component of Abell 1758, that would have not been detected on its own.}
\label{LXZ}
\end{figure}

The reason that upper limits are a key problem in looking at correlations is the virtual degeneracy between luminosity and redshift in X-ray samples, as shown in Figure~\ref{LXZ} \citep[see also][]{ebe96,vent07}.  This means that any potential correlations and detection experiments must be given very careful scrutiny to avoid redshift dependent observational biases. This  is the motivation behind the current work.  There have been a variety of indirect arguments given in the literature  to show that selection effects are not likely to be responsible for the observed correlations;  these will be discussed in light of the current results in Section \ref{compare}. A complementary survey to ours by \cite{mark04} using  VLA data also includes upper limits, but has not been published. 

\section{ Analysis }

\subsection{Sample Definition}

The sample of sources studied here was selected from the ROSAT Brightest Cluster Sample (hereinafter, BCS) of \cite{ebe98}.  That sample was extended beyond the earlier Abell cluster, XBACS, sample \citep{ebe96} by including Zwicky clusters and clusters selected on the basis of their X-ray emission alone.  The BCS  is estimated to be 90\% complete for z$<$0.3, and contains 201 northern latitude ($\delta \geq$0$^\circ$) clusters with $\vert$b$\vert \geq$20$^\circ$ and 0.1-2.4 keV fluxes higher than 4.4$\times$10$^{-12}$ erg~cm$^{-2}$~s$^{-1}$. 
We  further selected only those clusters with $\delta >$30$^\circ$, in order to overlap with the Westerbork Northern Sky Survey \citep[WENSS,][]{wenss} at 327 MHz. 

\begin{figure}[h]
\begin{center}
\includegraphics[height=8.5cm]{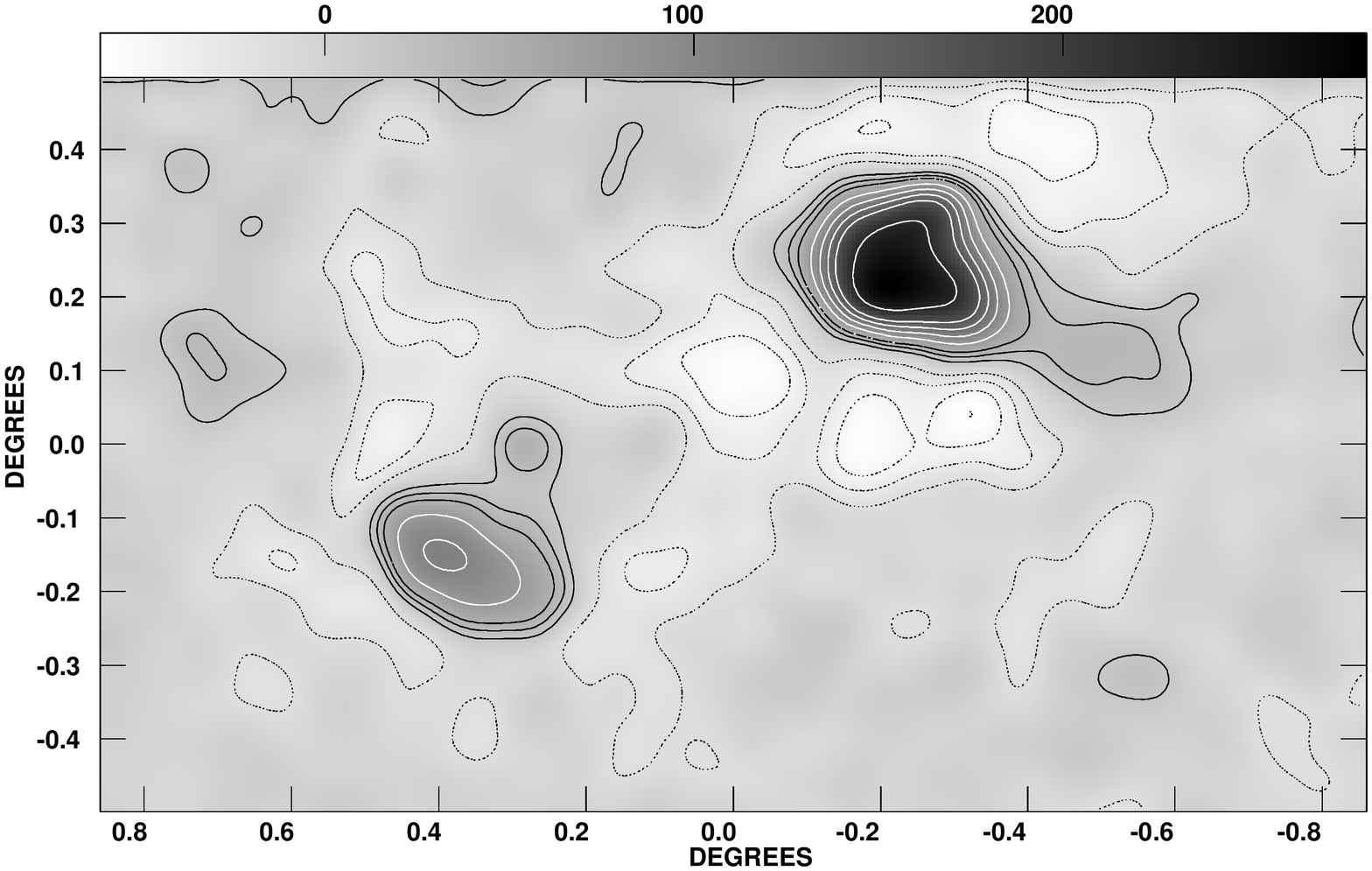}
\vskip 0.2in
\includegraphics[height=8.5cm]{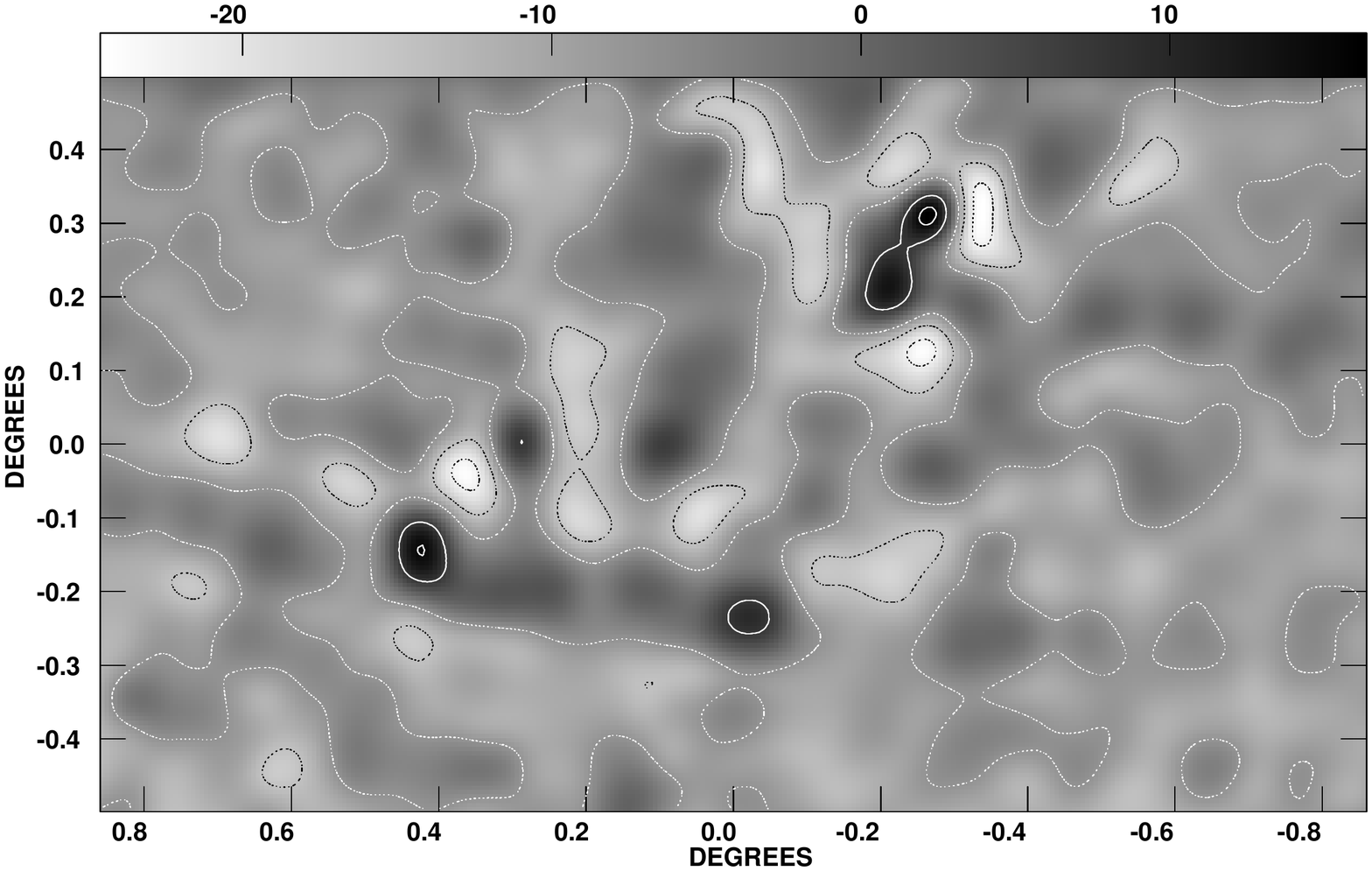}
\end{center}
\caption{Comparison of the NVSS and WENSS sensitivities to low surface brightness emission. The grey scale shows the large scale structure in NGC~6251, convolved to a resolution of 240'', from the WENSS (top) and NVSS (bottom) surveys.  Small scale emission was first removed using the filtering procedure described in the analysis section.  Contour levels are 0.015 mJy/beam (WENSS) and 0,075 mJy/beam (NVSS) $\times$ [-3, -2, -1, 1, 2, 3, 5, 7, 9, 11].}
\label{WvN}
\end{figure}

The WENSS has an angular resolution of 54''$\times$54''csc($\delta$), and an rms scatter of 3.6~mJy/beam, $\approx$ 5 times above the rms brightness fluctuations of the NRAO Very Large Array Sky Survey \citep[NVSS,][]{nvss} of 0.45~mJy in a 45'' beam at 1.4 GHz. Thus, for diffuse emission on scales of $\approx$100s of arcseconds, the NVSS is more sensitive unless the spectral index of the source is steeper than -1.1 \citep[ e.g., -1.7, as in A1914,][]{bacch03}.  However, at redshifts z$<$0.06, where 1~Mpc corresponds to  $>$900'', the NVSS sensitivity essentially disappears. There has been no evaluation, to our knowledge, of how the brightness sensitivity of the NVSS actually declines as the angular size increases towards the limit.  Figure \ref{WvN} illustrates this loss in sensitivity with a comparison between NVSS and WENSS of the diffuse structure of the giant radio galaxy NGC6251.

In order to facilitate an automated measurement procedure, we  eliminated a  number of sources that were too nearby (z$<$0.02) or too close to the declination limits of the WENSS survey or had other problems including confusing sidelobe structures from nearby very strong sources.  After the above selections, the final sample contains 79 clusters.  Throughout, we use a concordance cosmology, with H=70 km/s/Mpc, $\Omega_{M}$=0.27 and $\Omega_{vac}$=0.73 (but see the qualifying note to Table 1).  A preliminary report on this work was presented in \cite{rudn06}.

\subsection{Analysis Procedure}

Maps from the WENSS survey were extracted via NASA's Skyview site \footnote { A service of the  Astrophysics Science Division  at NASA/  GSFC
and the High Energy Astrophysics Division of the Smithsonian Astrophysical Observatory (SAO)}, centered on the X-ray positions given in \cite{ebe98}.  Each field was 600 pixels on a side (21.245''/pixel, field size 3.54$^\circ$), independent of redshift. We first show the {\it total} WENSS  flux within a circular aperture of 1~Mpc diameter centered on the X-ray emission (Figure \ref{total}) as a function of redshift, after subtracting the mean flux contribution averaged over the 3.5$^\circ$ field. The total flux is seen to be independent of redshift, as expected from a flux limited survey.  On this figure, we also show the detection limits for our diffuse flux search, as discussed below.  Note that the diffuse flux limits are typically  1-10\% of the maximum total flux, depending on redshift.  Thus, any automatic procedure to measure diffuse flux has to carefully remove the contributions from more compact radio sources in the direction of the cluster.

 In order to avoid the many uncertainties associated with visual searches, we developed a procedure to analyze these images requiring a minimum of subjective decisions. The essential elements of this procedure, described in more detail below, were:
\begin{enumerate}
\item  {\it Definition}  of the search region diameter as 1~Mpc, corresponding to angular diameters from ~220''~-~1660'' for 0.03$<$z$<$0.3 .  
\item Removal of compact sources of emission through automatic filtering of images.
\item Measurement of the residual diffuse flux in the search region, relative to a local background.
\item Identical analysis of a matched control region for each object, leading to a determination of flux detection limits as a function of redshift.
\item Simulation of halos as smooth Gaussian of 1~Mpc diameter, to assess detectability in our analysis procedure, and determine correction factors for the loss of smooth diffuse flux through filtering.
\item Preliminary morphological classification into halo, relic and radio-galaxy related diffuse sources.

\end{enumerate}

\subsection{Separation of diffuse from compact emission}

Our overall goal is to separate the diffuse flux associated with a cluster from the flux contributions of more compact sources, in a well-defined, objective manner.  We determined the diffuse flux within the 1~Mpc  apertures by filtering out the compact emission following the algorithm described in \cite{filter}. We used a filter size of $\approx$ 3~$\times$ the (elliptical) beam size for each declination (the filter is restricted to an odd number of pixels in each dimension).  Unresolved  sources are  reduced in flux by a factor of $\approx$10$^7$ by the filtering procedure, and therefore do not contaminate the diffuse flux.  The mean flux in the residual map (3.5$^\circ$ on a side) was then set to zero, manually excluding surrounding regions of very bright, diffuse emission when necessary. The flux within the 1~Mpc aperture was then summed. Setting the zero level as we did would reduce the flux for a 1~Mpc source by 1.5\% (0.03\%) at a redshift of 0.03 (0.3).  These are much smaller than other sources of error, especially the rms scatter in fluxes observed for blank control fields as discussed below, so no correction was made.  

The filtering procedure produces two additional effects.  First, it can remove flux from true halos, changing their detectability, as discussed below.  Second, sources that are somewhat extended, but less than 1~Mpc in size, will contribute to, and confuse the halo measurements.  We therefore carried out a suite of simulations to address both of these issues. as discussed in Section \ref{sim}.

For each cluster, we determine a net (filtered) diffuse flux above the local background in the 1~Mpc aperture;  this number can be either positive or negative.   Figure \ref{A980} illustrates the filtering procedure for one cluster in our sample, Abell 980.  The original WENSS image is shown at the top, with the middle panel showing the diffuse emission in greyscale, overlaid by contours of the original WENSS image.  Although the compact emission has been removed, a significant amount of related diffuse emission is still present.  The boxy shape of the diffuse flux is an artifact of the filtering procedure; the smallest visible boxes show the filter size. The diffuse emission is overlaid in the bottom panel  with smoothed contours from the ROSAT broadband image.  In this particular case, because there was extended emission visible surrounding the compact sources, we classified it as ``radio galaxy'' emission; we do not claim a detection for a halo or relic in this cluster, although further study would be useful.  We note that mini-halos surrounding a central radio galaxy would be classified as ``radio galaxy'' according to our procedure.

\subsection{Halo Simulations} \label{sim}

We simulated the detectability of Gaussian halos by injecting such halos into a random patch of the WENSS survey with no obvious sources and carrying out the above filtering procedure, as a function of both redshift and declination (which change the filter size).  The filtering has the effect of reducing the flux.   The results of this analysis are shown in Figure \ref{diam} for each field, plotted in terms of the diameter of a halo at which 50\% of the flux would remain in the residual, diffuse image and 50\% would have been filtered away.  Thus, at a redshift of 0.1, for example, a 500~kpc diameter halo would yield only 50\% of its flux in our filtered, diffuse flux image.   Larger diameter halos would yield larger diffuse fluxes, and vice versa.  

A 1~Mpc diameter halo of a given true flux would therefore have a different detectability at different redshifts (and declinations).  At each redshift the \emph{filtered} halo flux (not the true halo flux) must rise above the threshold flux determined from the control fields, as discussed below, to be considered a detection.  This is a strict quantitative criterion, not relying on visual inspection, and subject only to further classification of the type of emission detected above the threshold flux.

Once a halo has been detected, we need to correct for the results of the filtering to calculate its true flux and luminosity.  We therefore determined ``correction factors'' to be applied to each detection, so that they are all in terms of the equivalent flux of a 1~Mpc diameter Gaussian halo (true, before filtering).  These correction factors are shown in Figure \ref{diam}.

\subsection{Detection Threshold}
Having simulated the appearance of 1~Mpc Gaussian halos in our experiment, we must now determine when a given measurement constitutes a detection.   The normal procedure for this is to determine the rms noise ($\sigma$) in the measurements, and demand, e.g., that the signal be  $>$n$\times\sigma$.  However, in the current analysis, such a procedure is not reliable, because the amount of noise power as a function of angular scale is not known {\em a priori}.  In addition, the  ``noise'' itself contains instrumental contributions (e.g., grating lobes and other large scale background variations), contributions from residual flux associated with background radio sources, etc.

We therefore adopted a different procedure, which automatically includes all sources of errors.  We set up a group of control fields, and conducted the identical analysis on them as we did for our cluster sample.  Because the beam size and filter size are a function of declination, we therefore set up a a control field 2$^\circ$ north or south of each cluster (including several clusters later eliminated).  We then assigned that control field the same redshift as the original source in order to fix the size of the 1~Mpc aperture.  We determined the residual diffuse fluxes (positive or negative) for each control field, and plot these as a function of redshift in Figure \ref{control}.   Cluster fluxes would then be considered detections only if they were sufficiently above the distribution of control fluxes, in a manner we now describe.

 At each redshift, the diffuse fluxes from the control fields had a {\em mean} $\approx$~0, as expected if they did not contain true diffuse emission.  However, the {\em rms scatter} in the diffuse control fluxes fell approximately linearly with (assumed) redshift, which can also be seen in Figure \ref{control}.  This rms decrease as a function of redshift is expected, since the angular size of the 1~Mpc aperture is decreasing.   If the noise in the filtered WENSS survey had an approximately flat power spectrum,  then the number of independent beams within the aperture would fall $\approx$ 1/z$^2$, and the rms noise as $\approx$1/z.  

For the cosmology we used, the number of independent beams actually falls more slowly than 
$\approx$ 1/z$^2$, and we could have adopted a different detection threshold dependence.  However, given that the noise power spectrum of the filtered WENSS is unknown, except for the control fields studied here, we decided to use the simplest empirical (1/z) trend.
We therefore adopted a detection threshold of (7.5/z)~mJy, which is shown in Figure \ref{control}, up to a limit of z=0.25.  This threshold would allow only 3 out of 85 ``false'' detections (a 96\% confidence limit) from the control sample.  If the diffuse flux from a cluster was above this threshold, it was considered a detection.  In Figure \ref{distribute} we plot the distribution of the (diffuse flux/threshold flux) for both control and cluster samples.  This shows a large population of cluster sources above the detection threshold, while admitting only 3 (4\%) of the controls. 

The situation above z=0.25 needs further explanation.  Four of the raw, uncorrected cluster measurements are far above those of the control fields (Figure \ref{control}), although three of them would fall below the extrapolation of the 1/z detection threshold discussed above, and not be counted as detections.  However, since our filtering procedure begins to significantly reduce the observed fluxes (both cluster and control) above z=0.25 (Figure \ref{diam}), an extrapolation of the 1/z trend may not be appropriate.   We therefore asked whether the clusters and controls were separable statistically above z=0.25, without regard to a pre-determined detection level.  Both contingency table and  rank-sum tests yield a probability of 4\%-6\% of achieving a source/control separation as strong as observed, so the four high flux clusters can be considered detected with a 94-6\% confidence. The one cluster with a very low flux is Abell 1758b, which is confused by the nearby presence of Abell~1758a, and was later removed from the sample.  We therefore consider the four clusters well-separated from the control fields as detections, although we will be explicit whenever we include them in later statistical arguments. We note that these four sources have already been detected in other studies, as detailed in Table 1.

\subsection{Contributions from extended sources}

This same simulation analysis described above can be used to determine the possible contamination from slightly extended sources in each field.  While unresolved sources are reduced in flux by 10$^7$ in the diffuse residual image,  a 200~kpc radio galaxy at redshift 0.05 would, according to Figure \ref{diam} contribute half its flux to the diffuse image.  This contamination is unavoidable, since there is no way for the analysis to isolate the origins of the diffuse flux.  We therefore accept the total diffuse flux numbers as measured, and, after the fact, attempt to visually assess whether radio galaxy contamination is likely or not.   This procedure differs from manually determined halo fluxes, where a subjective decision of how much diffuse emission belongs to a halo and how much belongs to individual radio galaxies must sometimes be made.

\section{Results}

The sources detected using these criteria are listed in Table 1.  We visually inspected each detection and  classified them as follows:  a) Radio galaxy - diffuse emission  likely associated with somewhat extended  source(s) visible in the original WENSS image;  b) Halo - diffuse emission centrally located with respect to the X-ray emission;  c) Relic - diffuse emission not centered on the X-ray emission.   These classifications should all be considered tentative pending deep images of sufficient resolution. Since most of the detections already have such images published,  the suggested identification is confirmed in many cases. In the case of the radio galaxy class, it is possible that deep images would also show diffuse emission not associated with the compact sources, but we cannot isolate that from the WENSS images alone. We also note that our detection scheme is biased against finding relics at distances greater than 500 kpc from the X-ray centroid, and against finding sources with a large fraction of their flux in small-scale substructures.   Table 2 lists those sources that do not meet our detection criteria.  The upper limits listed are set at the detection threshold at each redshift.

In Figure \ref{lit}, we plot the fluxes from our measurements as a function of those estimated from the literature, using a spectral index of -1.2 to convert from 1.4~GHz fluxes as needed.  We preferentially find larger fluxes than do \cite{kemp01}, as expected since they restricted themselves to only counting flux above a 2$\sigma$ threshold.  We preferentially find smaller fluxes (assuming the spectral index is correct) than \cite{cass07}, as expected since we are sensitive only to the smooth component of halo emission, and we automatically filter out any contamination from smaller scale structure due to extended radio galaxies as well.  This comparison has two noteworthy consequences.  First, our method produces reasonable results in comparison with other measurements.  Second,  the specific method used to determine diffuse cluster fluxes has a significant effect on the measurements.

A number of our detected diffuse sources are either new, or provide additional information to what is already known.  For example, we find a new Mpc-scale relic at the edge of the X-ray cluster RXJ1053.7+5450, a new halo and possible relic associated with Abell 2061,  a serendipitous diffuse X-ray source near poor clusters in the Abell~781 field, and confirm extended emission outside of Abell 2255.  Brief notes and figures on sources with significant new information are given in the Appendix.

In the tables and plots, radio luminosities do not include any k-corrections.  The X-ray luminosities are taken from \cite{ebe98} converted from their Euclidean cosmology with H=50~km/s/Mpc to the concordance one.  Figure \ref{XvR} plots the diffuse radio vs. X-ray luminosity, including both detections and upper limits.  At a fixed X-ray luminosity, the detections and upper limits for radio luminosity overlap because of the (small) spread in redshift at each X-ray luminosity.

Although the plot in Figure \ref{XvR} represents the actual observations, it is a biased estimate of the diffuse flux, due to the effects of our filtering procedure. Referring back to Figure \ref{diam}, we see that at z=0.25~-~0.3, for example, only 50\% of the diffuse flux would be observed for a  source with a diameter of 1~Mpc.  We therefore applied the correction values for 1~Mpc halos as a function of redshift and declination as described earlier. Corrected values for flux and luminosity are given in Tables 1 and 2, and also plotted in Figure \ref{XvR}.  In most cases, the corrections are quite small, but become very large beyond z$>$0.25, as can be seen at the bottom of Figure \ref{diam}.   When the corrections are as large as a factor of 2,  they are also quite uncertain because they depend on the (unknown) shape of the diffuse emission.  

There is a second, more important loss of diffuse flux for which we cannot correct.  If a halo or radio galaxy contains substructure on scales comparable to those in Figure \ref{diam}, then the substructure will also be partially removed along with the compact source emission. In this regard, the current survey has the opposite biases to those done by visual inspection of interferometer images, where patchy emission can {\it increase} the visibility of the halo, especially at low redshifts.   Since few halos are completely smooth \citep[see, e.g.,][]{gov01}, we expect that our halo fluxes will be lower than those determined by visual inspection where small contaminating sources can be removed by hand, and everything larger is assigned to the halo. Our preferentially lower values for halo emission from the brightest X-ray clusters (highest redshifts) can be seen in  Figure \ref{XvR} when compared with the radio-X-ray luminosity trend taken from \cite{cass07} and scaled to 330 MHz a spectral index of -1.2. Our diffuse radio galaxy emission may be even more strongly reduced by filtering, since it is not well modeled by a smooth Gaussian. Our measurements should thus be understood as due to the ``smooth'' component alone, i.e., on scales larger than shown in Figure \ref{diam}.  However, this highlights an even more subtle problem;  when a visual inspection is made to separate radio galaxy and halo emission, it is not clear how to do this in a meaningful way on scales of $\ge$100~kpc.

\section{Discussion}

Our survey was designed to isolate Mpc-scale  diffuse emission in clusters, independent of its origin.  We begin our discussion with those findings, and then make some specific comments about radio halos.

Examination of Figure \ref{XvR}  reveals a fairly well-defined upper envelope to the radio luminosity of diffuse structures in clusters of fixed X-ray luminosity.  There is no observational constraint that would have prevented powerful diffuse radio emission (e.g., $>$10$^{24.5}$~W/Hz) from being detected in clusters with X-ray luminosities $\approx$10$^{44}$~erg/s.   Above  L$_X$=10$^{44.95}$erg/s (chosen to make the best case), we detect 4 out of 7 [8 out of 11, including z$>$0.25] clusters with some form of diffuse emission.  For $L_X<10^{44.95}$, however, we detect only 13 out of 68 clusters, despite the fact that our radio detection limits are approximately a factor of three lower at the lower X-ray luminosities. Using a two-by-two contingency table, these detection rates are different at the 0.02 [10$^{-4}$] level.  Another indicator of this correlation, although not quite as significant, is that the X-ray luminosities of clusters with diffuse radio emission are higher than those without detected diffuse emission at the 0.12 [0.01] confidence level, using the Wilcoxon two-sample test.  

Looking specifically at the halos, we have an overall detection rate of 7  out of 74, or 9-11\%.  Including the sources  with z$>$0.25  yields 9 out of 79, or 11\%.  At the highest X-ray luminosities, L$_X > 10^{44.95}$~erg/s,  the halo detection rate is  $\sim$50\% (5 out of 11, counting z$>$0.25), similar to the values cited by \cite{gioferr01} and GCLUS.

The correlation of  radio and X-ray luminosities likely reflects some underlying limit to the energy available to the diffuse relativistic particles and fields in each cluster.  The exact functional form of this correlation is, however, unclear, due to the large number of non-detections.   Weak halos such as found in Abell~2256 \citep{hr2256} could easily populate the lower right of Figure \ref{XvR} either because they are too weak or confused by a strong relic or radio galaxy.   Whether a halo is detected, and with what flux, depends in a complicated way on the exact observational parameters and analysis.

From this study of X-ray selected clusters at 0.03$<$z$<$0.3,  we can neither  confirm nor rule out the hypothesis that at each X-ray luminosity, the radio luminosity takes on a wide range of values up to the upper limit. The alternative hypothesis, similarly unresolved in this analysis,  is that there is a bi-modal distribution of radio luminosities at each X-ray luminosity, i.e., with the high radio luminosities along the trend line, and a second set of ``off'' clusters with much lower radio luminosities.  This distinction is key for some physical models of halos, as discussed below. 

While the above considerations suggest confirmation of the strong relationship between maximum radio halo luminosity and X-ray emission, there are two important caveats that could change our understanding of this phenomenon.  The diffuse luminosity we have associated with radio galaxies (RG) is also well-correlated with the X-ray luminosity (Figure \ref{XvR}, $<$1\% probability level using a rank-sum test), driven largely by the redshift dependence.  Below L$_X$=10$^{44.5}$ erg/s, we detect only radio galaxy emission; above L$_X$=10$^{44.5}$ erg/s the RG diffuse emission populate the same portion of the (L$_X$, L$_R$)  diagram as the radio halos and relics.

There are two types of possible explanations for this.   The distribution in the (L$_X$, L$_R$) plane could be governed by a variety of selection effects.  Alternatively, at least some halos could simply be radio galaxy emission on scales of 10$^{2.5-3}$ kpc, where the association with the parent galaxy is no longer obvious.  This possibility is discussed briefly in Section \ref{physical}.
Second, there do exist diffuse cluster-like sources (halo or relic-type structures) with low X-ray luminosities \citep{delain06}, found only in {\em non}-Xray-selected surveys.  A detailed study of the 0809+390 system \citep{brown08}, with two likely independent diffuse radio patches, shows that at least one of them shares all the properties of a classic radio relic, but with little or no extended X-ray emission.  The optical environments also appear to be very different than those in X-ray selected samples.  At present, there is no way to cleanly incorporate these low X-ray luminosity systems into models of the acceleration of relativistic plasmas in rich clusters \citep{mini01,pfrom07b}

For the most luminous X-ray clusters, we have a high enough detection rate to look at the association of merger activity with diffuse radio emission.   Looking at the twelve highest X-ray luminosity clusters (including those with z$>$0.25), we find that for the eight detected cases of diffuse radio emission of all types, six clusters (665, 773, 781, 1682, 1758A, 1914) show disturbed or double-peaked X-ray morphologies \citep{bauer05,ota, barrena07}, while two appear regular (697, 1953).  However, the velocity structure of Abell~697 also shows evidence of a past complex merger \citep{girardi06} and no detailed optical work yet exists for Abell~1953.  Turning to the four {\it non-detections}, two appears disturbed (1763 and 2111 \cite{ota}) and three appear regular (2219, 2261, \cite{ota}), although \cite{boschin04} suggest evidence for a past merger in 2219. This small sample supports the commonly held view that clusters with diffuse  radio emission show evidence for merger activity;  however, this is also true for at least some of the non-radio-detected clusters as well.

\subsection{Comparisons to earlier work}
\label{compare}  

 Except for the newly published GCLUS work over the narrow redshift range 0.2$<$z$<$0.4, only a few  upper limits are  published for radio emission from Mpc-scale halos \citep{bacch03,ferr05}. Without upper limits, it is difficult to evaluate possible selection effects that could influence the observed radio halo luminosity vs. cluster X-ray luminosity relationship.    We examine those selection issues here, in the light of our results and recognizing the near degeneracy between X-ray luminosity and redshift in most samples. In general, the claim of a strong correlation is limited to radio  luminosities above 10$^{24.5}$ W/Hz or X-ray bolometric luminosities above 10$^{45}$ erg/s \citep{ferr00,bacch03}.

The first key assumption made in the literature is that in a survey such as the NVSS, or any interferometer observations,  there is a limiting surface brightness that is independent of the angular diameter up to the survey limit.\footnote{This constant brightness translates to different rest-frame brightnesses at different redshifts, but that is not germane to the current argument.}  Table 3 in \cite{gioferr00} explicitly gives surface brightness limits for their non-detected clusters which are meant to apply on all angular scales up to the maximum allowed by the interferometer observations.  However, as we show in Figure \ref{contours}, the detectability by eye of diffuse sources cannot be described by a size-independent surface brightness.  In this pair of contour plots we have inserted  circular Gaussian halos into an NVSS image at high galactic latitude.  Two different values for the Gaussian FWHM are used, 300'' (1~Mpc at z=0.2) and 900'' (1~Mpc at z=0.06), both with the same peak surface brightness of 0.7mJy/45''beam, or $\approx$1.3$\sigma$ for this field. The effect of an extended low signal:noise source is to bias the local noise so that more independent beams rise above the contouring threshold.  The larger size thus appears much  easier to detect; we expect that this will lead, for visual searches, to higher surface brightness detection limits for smaller sources and for sources at high redshifts. The same incorrect assumption about constant surface brightness sensitivity is used by \cite{clark05} in her calculation (not measurements) of the sensitivity limit of the NVSS as a function of redshift.  It is important for visual searches to avoid such constant surface brightness claims, and to use techniques such as the insertion of fake sources into images (as in GCLUS and here) to determine detectability.

There is an additional selection effect at the {\em low} redshift end.  The sensitivity of the NVSS (the most commonly used all-sky survey for halos) falls off sharply for sources with  angular sizes ($\approx$15', or 1~Mpc at z=0.06). Thus, the surface brightness of 1~Mpc halos below this redshift would have to be much higher than above this redshift, in order to be detected.   At the same time, as argued above, it is harder to visually detect smaller (higher redshift) halos of the same surface brightness.  There is thus likely to be a ``sweet spot'' in redshift for NVSS detections around z$\approx$0.1; detailed modeling of how more realistic  halo structures would be visually detected in the NVSS needs to be done to rule out both high and low redshift biases in any claimed correlations.

A brightness selection effect also occurs in our objective WENSS search. Figure \ref{sbright} shows the surface brightnesses for our detections and non-detections, as a function of X-ray luminosity.  Similar to the trends of radio luminosity with X-ray luminosity, we find high radio brightness sources only at high X-ray luminosities.  However, the upper limits show that we would be unable to detect low radio brightness sources in high X-ray luminosity clusters, so the detections alone provide a biased view of the correlation.

Recognizing that surface brightness biases occur in all  searches, can the selection effects in visual searches be isolated or estimated? \cite{ferr05} notes that her three plotted upper limits in radio brightness are along the same trend as the detections, but at lower X-ray luminosities. For surveys prior to GCLUS, it is not clear whether the same problem (upper limits in brightness showing the same trend as detections) exists. Observed radio brightnesses are often significantly above the (assumed size-independent) brightness limits in visual surveys \citep[e.g.,][]{cass06}, which could lead one to believe that the non-detections are of much lower brightness.  However, this requires a much more quantitative analysis.  

These caveats notwithstanding, there is useful information in our surface brightness distributions.  We see that  the  distribution of brightnesses cannot rise sharply towards lower brightnesses, or the detected sources would be peaked up at the survey limit.  We can therefore conclude that there is either a broad, fairly uniform distribution of surface brightnesses, or a bimodal distribution (on/off) significantly below our upper limits (as now claimed by GCLUS), but we cannot distinguish between these with the current sample.

We can also compare the slopes of two different correlations studied in both our objective survey and visual searches.  If the surface brightness values  in \cite{cass06} are converted to a fixed aperture of 1~Mpc diameter, there is a tight correlation between brightness and X-ray luminosity with $\frac{log(radio~ brightness)}{log(X-ray~ luminosity)} \approx$1.5 (R. Cassano \& G. Brunetti, private communication).  This is similar to the trend we see for our detections (except for a few radio galaxies with L$_X~<~$10$^{44.5}$ erg/s). However, since our limiting flux criterion forces our upper limits to follow the same trend,  we would therefore not claim an intrinsic correlation for our data.  Turning now to the luminosity correlation, above L$_X$=10$^{44.5}$erg/s our detections  follow a radio to X-ray luminosity slope similar to previously determined values,  value of 2 \citep{cass07}, or 1.8 \citep{ferr03}; again, our upper limits follow the same trend as the detections.  While we can conclude from our objective survey that there are no high brightness halos at low X-ray luminosities {\it for X-ray selected clusters},  we know very little about low brightness halos at high X-ray luminosities. Also, the data on the low radio luminosities of halos that could accompany low X-ray luminosity clusters is very sparse, and their detection and study will likely require the new generation of telescopes such as LOFAR, the Long Wavelength Array and the Square Kilometer Array.

In support of an intrinsic correlation between radio halo and X-ray luminosities, as opposed to selection biases, we can look to the strong correlation found by \cite{cass07},  $\frac{log(radio~luminosity)}{log (halo~radius)}~=~4.2~\pm~0.7$.  They conducted a variety of Monte Carlo type analyses (under the assumption of constant brightness sensitivity as a function of angular diameter)  and conclude that a selection driven slope would have been 2.5~$\pm$~0.4, different than what is observed.  In our survey and others, the limiting radio luminosity is higher at higher redshifts, and thus for higher X-ray luminosities.  Despite this, there is a higher fraction of detected diffuse radio sources at high X-ray luminosities, which argues for at least some intrinsic correlation.

We note again that there are other important selection biases in all searches for diffuse cluster emission.  In the current objective survey the biases are fairly well understood;  we are most sensitive to smooth structures $\approx$1~Mpc in diameter.  We have little sensitivity to structures located beyond 500~kpc from the X-ray centroid, although we have identified such cases by visual inspection of the diffuse flux images.  Also, we rapidly lose sensitivity for any diffuse structures with substantial flux at smaller scales.  A rough quantitative guide to this limitation is Figure \ref{diam} (top), which shows what size source loses half its flux, as a function of redshift.

\subsection{Notes on Physical Implications}
\label{physical}

Our first result, that we confirm the upper envelope to diffuse radio luminosity as a function of X-ray luminosity for X-ray selected clusters, implies that the latter provides an indicator of the total energy available for relativistic particle acceleration. For radio halos, for example,  ``re-acceleration'' scenarios \citep{brun01,petros01}, the underlying physical property could be cluster mass.   \cite{cass07} and \cite{gov01} show a strong correlation between halo radio power and cluster mass, although \cite{pfrom07} emphasizes that other factors may be more important in regulating the relativistic plasmas, such as how the magnetic energy density scales with cluster mass and the dynamical state of the cluster.  

Indeed, a combination of factors limits the lifetime of Mpc halos to $<$1 Gyr in re-acceleration models, including the shorter radiative lifetime of electrons, the drop in magnetic field strength and the decay of shocks and turbulence post-merger \citep{sar99, ensrot,cass06}.   The physics of this last factor, dissipation, is poorly constrained \citep[e.g.,][]{brun05}. This is why evidence for or against bi-modality of the radio luminosity is important; an  observed ``turn-off'' of radio emission at some time post-merger would be both a confirmation of the re-acceleration scenario and a unique diagnostic of the evolution of the magnetized thermal plasma. Alternative models in which energy is derived from cosmic ray protons would yield long-lived halos, because the loss times for protons are $\sim$10$^{10}$ years \citep{pfens,blasi99}.  However, this process may only be dominant in the central regions of the clusters \citep{pfrom07a,pfrom07b}, where there is often confusion from extended AGN emission. 

As noted earlier, the fact that Mpc-scale diffuse emission associated with radio galaxies is found in the same part of the L$_R$, L$_X$ diagram raises the question about whether these radio galaxy plasmas are also subject to re-energization processes in the cluster. Such re-energization is one mechanism to overcome the otherwise too-short lifetimes of synchrotron emitting electrons distributed over Mpc scales.  The same question arises in the case of mini-halos, which are often associated with central AGNs;  \cite{eile06} suggest that the AGN/cluster-wide energization transition occurs at scales of $\sim$100 kpc. 
We note again that in the current paper, an AGN/mini-halo system would be classified as ``radio galaxy''.

\begin{figure}[h]
\begin{center}
\includegraphics[height=16cm]{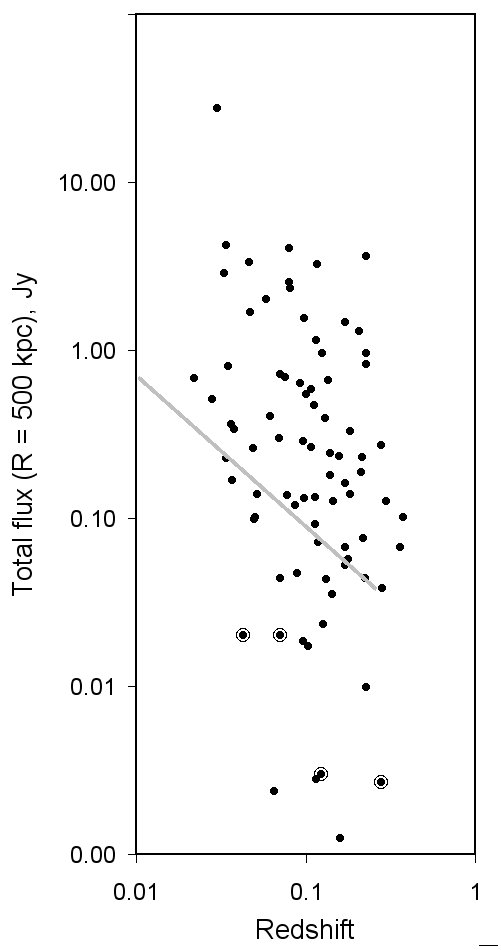}
\end{center}
\caption{Absolute value of the total flux within the R=500~kpc aperture centered on the X-ray centroid, as a function of redshift. Circled symbols indicate negative values.  To show the relationship with Mpc-scale diffuse emission, we indicate with the grey line the 96\% probability detection threshold for {\it diffuse} emission, as seen in Figure \ref{control} and discussed in the text.}
\label{total}
\end{figure}

An alternative explanation for diffuse radio galaxy emission scaling with X-ray luminosity is that the intrinsic power of radio galaxies could also be a function of cluster mass, perhaps indirectly related to the  mass available for accretion, for example. In that case, flows from the radio galaxy itself sustain the Mpc-scale emission.  This is apparently true for ``giant'' radio galaxies \citep{schoen00} in weak X-ray environments. However,  for the current cluster sample, if the intrinsic radio galaxy power scaled with cluster mass, we would also expect to see a correlation of X-ray luminosity  with the \emph{total} cluster radio luminosities, which includes the compact as well as diffuse emission. There instead appears to be a wide scatter in total luminosity, and little or no correlation. Figure \ref{total} shows, e.g., a large range of total radio fluxes at each redshift, while Figure \ref{control} shows only a narrow range of diffuse fluxes at each redshift, which leads to the L$_R$, L$_X$ correlation. 

\section{Concluding Remarks}
  \begin{itemize}
\item{We have provided an objectively defined set of measurements and well-defined upper limits to diffuse Mpc-scale radio emission in X-ray clusters and shown some of the interpretation pitfalls when such limits are not included.}
\item{Above X-ray luminosities of $\sim$10$^{45}$erg/s, most clusters in our sample (8 of 11) produce Mpc-scale diffuse radio emission in either halo or AGN-related form.}
\item{The apparent coincidence between the luminosities of halos and diffuse radio galaxy emission  is worth further study, and may point to more general properties of the energization of relativistic plasmas in clusters. It is also not yet clear whether a clean distinction can be drawn between patches of low brightness AGN-related emission and true halo emission. }
\item{A number of clusters studied here, and detailed in the Appendix, are worthy of further investigation to separate radio galaxy from halo emission, and to explore new relics and halos. The supercluster scale emission around Abell 2255 is especially important, and has recently been confirmed \citep{pizzo}.}
\item{Deep radio observations, with well-defined upper limits are needed for X-ray clusters over a broad range of redshifts  to extend and confirm the initial findings from GCLUS that halos may have a bimodal luminosity distribution. Similarly, surveys for diffuse radio sources in poor environments are necessary to test these relationships.}
\end{itemize}

\acknowledgments
   We gratefully acknowledge critiques and suggestions from G. Brunetti,  R. Cassano, G. Giovannini, C. Pfrommer and the anonymous referee that helped improve this paper considerably.  We thank K. Delain for help in setting up the WENSS analysis procedure. Partial support for this work at the University of Minnesota comes from the U.S. National Science Foundation grants AST~0307600 and AST~0607674.

\clearpage

\begin{deluxetable}{ccccccccccc}
\tabletypesize{\scriptsize}
\tablecolumns{11}
\tablecaption{Summary of detections\label{detections}}
\tablehead{
\colhead{Cluster} & \colhead{Morphology} & \colhead{RA}& \colhead{DEC}
& \colhead{z} & \colhead{Log(L$_X$)} & \colhead{Aper.} & \colhead{S$^{Tot}_{500}$} 
& \colhead{S$^{Diff}_{500}$} & \colhead{log(P$_R$)} & \colhead{Err} }
\startdata

       &   (reference)     & J2000& 	J2000&     &0.1-2.4 keV& ('') & (Jy)	& (Jy) &W/Hz &\\
	\hline
\setcounter{enumi}{1}
A665	&	\setcounter{enumi}{1} Halo~(\alph{enumi})
	&	127.739	&	65.85	&	0.1818	&	45.10	&	335	&	0.14	&	0.11	&	24.96	&	0.07	\\           						
	 & & & & & & & & (0.12) & (25.00) & \\
A773	&	\setcounter{enumi}{1}Halo~(\alph{enumi})
	&	139.475	&	51.72	&	0.217	&	45.01	&	291	&	0.08	&	0.04	&	24.67	&	0.20	\\
	& & & & & & & & (.05) & (24.77) & \\
A980	&	$\dagger$ RG+?
	&	155.617	&	50.12	&	0.1582	&	44.75	&	375	&	0.23	&	0.09	&	24.75	&	0.10	\\
	& & & & & & & & (0.11) & (24.75) & \\
A1033	&	\setcounter{enumi}{3}$\dagger$ RG?~(\alph{enumi})
	&	157.932	&	35.06	&	0.1259	&	44.59	&	449	&	0.95	&	0.11      &	24.61   &	0.11	\\
	& & & & & & & & (0.13) & (24.67) & \\
RXJ1053	&	$\dagger$ Relic 
	&	163.449	&	54.85	&    0.0704    &	43.89	&	764	&	0.72	&	0.36*	&	24.58	&	0.22	\\
	& & & & & & & & (0.37) & (24.58) & \\
A1132	&	\setcounter{enumi}{4}$\dagger$~RG~(\alph{enumi})
	&	164.616	&	56.78	&	0.1363	&	44.72	&	425	&	0.67	&	0.07      &	24.46   &	0.18	\\
	& & & & & & & & (.07) & (24.49) & \\
A1190	&	\setcounter{enumi}{5}RG~(\alph{enumi})
       &	167.869	&	40.83	&	0.0794	&	44.04	&	685	&	4.07	&	0.11       &       24.17	&	0.20	\\
       & & & & & & & & (0.11) & (24.18) & \\
A1314	&	\setcounter{enumi}{5}RG~(\alph{enumi})
	&	173.748	&	49.09	&	0.0338	&	43.29	&	1526	&	4.20	&	0.71	&	24.22	&	0.06	\\
	& & & & & & & & (0.71) & (24.22) & \\
A1682	&	\setcounter{enumi}{6} RG~(\alph{enumi},\setcounter{enumi}{10} \alph{enumi})
	&	196.739	&	46.55	&	0.226	&	44.95	&	357	&	0.96	&	0.07	&	24.97	&	0.10  \\
	& & & & & & & & (0.08) &(25.02) & \\
A1914	&	\setcounter{enumi}{1}Halo~(\alph{enumi})
	&	216.509	&	37.84	&	0.1712	&	45.15	&	352	&	1.48	&	0.05	&	24.52	&	0.22	\\
	& & & & & & & & (.06) & (24.63) & \\
A2034	&	\setcounter{enumi}{2}$\dagger$~Halo~(\alph{enumi})
	&	227.545	&	33.51	&	0.113	&	44.71	&	500	&	0.47	&	0.11	&	24.50	&	0.13	\\
	& & & & & & & & (0.13) & (24.56) & \\
A2061	&	$\dagger$ Halo 
	&	230.321	&	30.64	&	0.0777	&	44.47	&	698	&	(0.14)	&	0.27*	&	24.54	&	0.07	\\
	& & & & & & & & (0.29) & (24.57) & \\
A2061	&	\setcounter{enumi}{2}Relic~(\alph{enumi})
 	&	230.321	&	30.64	&	0.0777	&	44.47	&	698	&	(0.14)	&	0.12*	&	24.18	&	0.19	\\
	& & & & & & & & (0.13) & (24.21) & \\
A2218	&	\setcounter{enumi}{1}Halo~(\alph{enumi})
	&	248.97	&	66.21	&	0.171	&	44.86	&	352	&	0.05	&	0.05	&	24.51	&	0.22	\\
	& & & & & & & & (.05) & (24.55) & \\
A2256	&	\setcounter{enumi}{1}Relic~(\alph{enumi})
	&	256.01	&	78.63	&	0.0581	&	44.72	&	913	&	2.02	&	0.98*	&	24.84	&	0.05	\\
	& & & & & & & & (0.98) & (24.84) & \\
A2255	&	\setcounter{enumi}{1}$\dagger$ Halo~(\alph{enumi})
	&	258.182	&	64.06	&	0.0809	&	44.56	&	673	&	2.35	&	0.36	&	24.70	&	0.34	\\
	& & & & & & & & (0.36) & (24.71) & \\ 					
RXJ1733	&	\setcounter{enumi}{7} $\dagger$ RG+?~(\alph{enumi})
	&	263.255	&	43.76	&	0.033	&	43.59	&	1561	&	2.88	&	0.49	&	24.04	&	0.09	\\
	& & & & & & & & (0.50) & (24.04) & \\
Z8338	&	\setcounter{enumi}{8} RG~(\alph{enumi})
	&	272.71	&	49.92	&	0.0473	&	43.76	&	1108	&	1.67	&	0.35	&	24.20	&	0.37	\\ 
	& & & & & & & & (0.35) & (24.21) & \\ \hline \hline
A697	&	\setcounter{enumi}{2} Halo~(\alph{enumi})
	&	130.741	&	36.37	&	0.282	&	45.12	&	240	&	0.04	&	0.02	&	24.61	&	0.41	\\
	& & & & & & & & (.05) & (24.91) & \\	
Zw1953	&	\setcounter{enumi}{9} RG~(\alph{enumi})
	&	132.542	&	36.09	&	0.3737	&	45.45	&	198	&	0.10	&	0.02	&	24.86	&	0.31	\\
	& & & & & & & & (.05) & (25.33) & \\			
A781	&	\setcounter{enumi}{9}RG+?~(\alph{enumi})
	&	140.12	&	30.52	&	0.2984	&	45.14	&	230	&	0.13	&	0.02	&	24.67	&	0.37	\\
	& & & & & & & & (.07) & (25.25) & \\ 
A1758a	&       \setcounter{enumi}{2} Halo~(\alph{enumi})
	&	203.189	&	50.55	&	0.28	&	44.97	&	241	&	0.27	&	0.07	&	25.16	&	0.08	\\ 
	& & & & & & & & (.09) & (25.31) & \\	\hline

\enddata
\end{deluxetable}

\noindent Notes to Table 1:~ ~ \\
 Full names of abbreviated sources:  RXJ1053.7+5450; RXJ1733.0+4345.  \\Zw1953~=~ZwCl~0847.2+3617. Zw8338~=~ZwCl~1810.2+4949 

S$^{Tot}_{500}$ (S$^{Diff}_{500}$) refers to the total (diffuse) flux within a 500 kpc radius. P$_R$ is the monochromatic radio luminosity at 0.3 GHz.

Sources below the double line are from z$>$0.25 which fall below the nominal extension of the 96\% detection threshold in Figure \ref{control}, whose detection status is discussed in the text.

$\dagger$ new or extended detections of diffuse emission

 $ ^{\ast}$ apertures adjusted manually because of nearby  emission.   For 
Abell 2061, total fluxes not meaningful because of extensive nearby emission. For Abell 2256, we have classified the detection as a ``relic'', although we know from deep images that there is a small halo contribution as well \citep{hr2256}.

Total fluxes given in parentheses are not significant detections.

Corrected fluxes and radio luminosities are listed in parentheses when correction is $>$5\%.

The error in log radio luminosity is an approximation to the statistical error alone, calculated by assuming an uncertainty in log(flux) equal to one half the detection threshold at that redshift. It thus depends on both the redshift and how far the measured flux was above the threshold. Other important uncertainties and biases with these measurements are discussed in the text.

The angular size of the apertures in the table were the ones actually used, although the effective apertures are slightly bigger due to the finite size of the WENSS beam.  In order to partially compensate for this, the quoted apertures were calculated using a Hubble parameter of 72~km/s, rather than the 70~km/s used for the luminosity calculations. This $\sim$3\% adjustment, in retrospect, was much less than the uncertainties in the measurement.

\noindent References from column 2:\\
\setcounter{enumi}{0}
\addtocounter{enumi}{1} \alph{enumi}. \cite{giov99}; 
\addtocounter{enumi}{1} \alph{enumi}. \cite{kemp01};
\addtocounter{enumi}{1} \alph{enumi}. \cite{rola85}; 
\addtocounter{enumi}{1} \alph{enumi}. \cite{odea85}; 
\addtocounter{enumi}{1} \alph{enumi}. \cite{vall87}; 
\addtocounter{enumi}{1} \alph{enumi}. \cite{morr03}; 
\addtocounter{enumi}{1} \alph{enumi}. \cite{baue00}; 
\addtocounter{enumi}{1} \alph{enumi}. \cite{harr77};
\addtocounter{enumi}{1} \alph{enumi}. \cite{coor98};
\addtocounter{enumi}{1} \alph{enumi} \cite{vent08}.

\begin{deluxetable}{ccccccccc}
\tabletypesize{\scriptsize}
\tablecolumns{9}
\tablecaption{Non-detections of diffuse emission\label{non-detections}}
\tablehead{
\colhead{Cluster}  & \colhead{RA}& \colhead{DEC}
& \colhead{z} & \colhead{Log(L$_X$)} & \colhead{Aper.} & \colhead{S$^{Tot}_{500}$} 
& \colhead{S$^{Diff}_{500}$} & \colhead{log(P$_R$)} }
\startdata
        &  J2000    &   J2000     & & 0.1-2.4keV&    ('')   & [Jy]&  [Jy]& W/Hz \\ \hline	
A7			&	2.946	&	32.42	&	0.1073	&	44.58	&	523	&	0.27	&	$<$0.07	&	$<$24.26		\\		
A77			&	10.123	&	29.55	&	0.0712	&	44.10	&	756	&	(0.04)	&	$<$0.11	&	$<$24.06	\\		
A272			&	28.766	&	33.90	&	0.0872	&	44.39	&	629	&	0.12	&	$<$0.09	&	$<$24.16		\\		
RXJ0228		       &	37.069	&	28.18	&	0.035	&	43.28	&	1475	&	0.80	&	$<$0.22	&	$<$23.73		\\
A376			&	41.534	&	36.89	&	0.0488	&	44.00	&	1075	&	0.26	&	$<$0.16	&	$<$23.88		\\
A407			&	45.456	&	35.84	&	0.0464	&	43.56	&	1128	&	3.33	&	$<$0.16	&	$<$23.86	\\
A566			&	106.093	&	63.28	&	0.098	&	44.55	&	567	&	1.54	&	$<$0.08	&	$<$24.21		\\
A576			&	110.382	&	55.76	&	0.0381	&	44.00	&	1361	&	0.34	&	$<$0.20	&	$<$23.86		\\
A586			&	113.093	&	31.63	&	0.171	&	44.93	&	352	&	0.07	&	$<$0.04	&	$<$24.50		\\
UGC03957		&	115.239	&	55.44	&	0.0341	&	43.83	&	1513	&	0.23	&	$<$0.22	&	$<$23.98		\\
RXJ0751		       &	117.842	&	50.21	&	0.022	&	43.18	&	2312	&	0.68	&	$<$0.34	&	$<$23.52	\\
A602			&	118.351	&	29.37	&	0.0621	&	43.92	&	858	&	0.40	&	$<$0.12	&	$<$24.00		\\
Z1478			&	119.919	&	54.00	&	0.1038	&	44.25	&	538	&	(0.02)	&	$<$0.07	&	$<$24.24	\\
RXJ0819	        	&	124.913	&	63.61	&	0.119	&	44.33	&	381	&	0.07	&	$<$0.06	&	$<$24.31	\\
A646			&	125.547	&	47.10	&	0.1303	&	44.54	&	442	&	0.39	&	$<$0.06	&	$<$24.36		\\
A655			&	126.361	&	47.13	&	0.1267	&	44.71	&	452	&	(0.02)	&	$<$0.06	&	$<$24.34	\\
A667			&	127.019	&	44.76	&	0.145	&	44.55	&	403	&	0.13	&	$<$0.05	&	$<$24.41	\\
A671			&	127.17	&	30.43	&	0.0503	&	43.82	&	1044	&      (0.10)	&	$<$0.15	&	$<$23.90		\\
A757			&	138.357	&	47.69	&	0.0514	&	43.82	&	1024	&	(0.14)	&	$<$0.15	&	$<$23.91		\\
Z2701			&	148.198	&	51.89	&	0.214	&	44.92	&	294	&	0.23	&	$<$0.04	&	$<$24.62	\\
Z2844			&	150.657	&	32.69	&	0.05	&	43.62	&	1051	&	(0.10)	&	$<$0.15	&	$<$23.90		\\
A961			&	154.085	&	33.64	&	0.1241	&	44.38	&	461	&	(0.00)	&	$<$0.06	&	$<$24.33	\\
A963			&	154.255	&	39.03	&	0.206	&	44.91	&	303	&	1.29	&	$<$0.04	&	$<$24.60	\\
A990			&	155.912	&	49.15	&	0.144	&	44.77	&	406	&	(0.04)	&	$<$0.05	&	$<$24.41		\\
A1068			&	160.187	&	39.95	&	0.1386	&	44.77	&	419	&	0.18	&	$<$0.05	&	$<$24.39		\\
A1302			&	173.307	&	66.40	&	0.116	&	44.36	&	488	&	(0.00)	&	$<$0.07	&	$<$24.30		\\
A1361			&	175.917	&	46.37	&	0.1167	&	44.43	&	486	&	3.25	&	$<$0.07	&	$<$24.30	\\
A1366			&	176.202	&	67.41	&	0.1159	&	44.47	&	489	&	1.15	&	$<$0.07	&	$<$24.30		\\
A1423			&	179.342	&	33.63	&	0.213	&	44.90	&	295	&	0.19	&	$<$0.04	&	$<$24.61	\\


%
RXJ1205	&	181.299	&	39.34	&	0.037	&	43.44	&	1399	&	(-0.17)	&	$<$0.21	&	$<$23.76	\\
A1677	&	196.471	&	30.89	&	0.1832	&	44.69	&	333	&	0.33	&	$<$0.04	&	$<$24.53	\\
RXJ1320	&	200.035	&	33.14	&	0.0362	&	43.28	&	1428	&	0.36	&	$<$0.21	&	$<$23.75	\\
A1763	&	203.818	&	41	&	0.2279	&	45.07	&	280	&	3.65	&	$<$0.03	&	$<$24.66	\\
A1767	&	204.032	&	59.21	&	0.0701	&	44.26	&	767	&	0.3	&	$<$0.11	&	$<$24.06	\\
A1885	&	213.432	&	43.66	&	0.089	&	44.25	&	618	&	-0.05	&	$<$0.09	&	$<$24.17	\\
Z6718	&	215.401	&	49.54	&	0.071	&	43.96	&	758	&	(-0.02)	&	$<$0.11	&	$<$24.06	\\
A1902	&	215.428	&	37.3	&	0.16	&	44.63	&	372	&	(0.00)	&	$<$0.05	&	$<$24.46	\\
A1918	&	216.342	&	63.18	&	0.1394	&	44.50	&	417	&	0.25	&	$<$0.05	&	$<$24.39	\\
A1930	&	218.12	&	31.63	&	0.1313	&	44.49	&	439	&	(-0.04)	&	$<$0.06	&	$<$24.36	\\
A2064	&	230.221	&	48.67	&	0.1076	&	44.40	&	522	&	0.59	&	$<$0.07	&	$<$24.26	\\
A2069	&	231.041	&	29.92	&	0.1145	&	44.83	&	494	&	0.09	&	$<$0.07	&	$<$24.29	\\
A2111	&	234.924	&	34.42	&	0.229	&	44.94	&	279	&	(-0.01)	&	$<$0.03	&	$<$24.66	\\
A2110	&	234.952	&	30.72	&	0.098	&	44.47	&	567	&	0.13	&	$<$0.08	&	$<$24.21	\\
A2124	&	236.25	&	36.07	&	0.0654	&	44.00	&	818	&	(0.00)	&	$<$0.12	&	$<$24.02	\\
A2175	&	245.128	&	29.89	&	0.0972	&	44.34	&	571	&	-0.02	&	$<$0.08	&	$<$24.21	\\
A2199	&	247.165	&	39.55	&	0.0299	&	44.43	&	1717	&	27.7	&	$<$0.25	&	$<$23.67	\\
A2219	&	250.094	&	46.71	&	0.2281	&	45.21	&	280	&	0.82	&	$<$0.03	&	$<$24.66	\\
A2241	&	254.934	&	32.62	&	0.1013	&	44.16	&	550	&	0.55	&	$<$0.07	&	$<$24.23	\\
A2244	&	255.667	&	34.06	&	0.097	&	44.84	&	572	&	0.29	&	$<$0.08	&	$<$24.21	\\
A2249	&	257.452	&	34.47	&	0.0802	&	44.47	&	679	&	2.56	&	$<$0.09	&	$<$24.11	\\
RXJ1715	&	258.826	&	57.43	&	0.028	&	43.49	&	1829	&	0.51	&	$<$0.27	&	$<$23.64	\\
Z8197	&	259.548	&	56.67	&	0.1135	&	44.34	&	498	&	0.13	&	$<$0.07	&	$<$24.28	\\
A2261	&	260.615	&	32.13	&	0.224	&	45.16	&	284	&	0.04	&	$<$0.03	&	$<$24.64	\\
A2294	&	260.98	&	85.89	&	0.178	&	44.71	&	341	&	0.06	&	$<$0.04	&	$<$24.52	\\
RXJ1740	&	265.133	&	35.65	&	0.043	&	43.56	&	1213	&	(-0.02)	&	$<$0.18	&	$<$23.82	\\
Z8276	&	266.056	&	32.98	&	0.0757	&	44.47	&	715	&	0.69	&	$<$0.10	&	$<$24.09	\\
RXJ1750	&	267.567	&	35.08	&	0.171	&	44.63	&	352	&	0.16	&	$<$0.04	&	$<$24.49	\\
A2312	&	283.451	&	68.39	&	0.0928	&	44.15	&	595	&	0.64	&	$<$0.08	&	$<$24.19	\\
\enddata
\end{deluxetable}
\clearpage
\noindent Notes to Table 2:~ ~ \\
\vskip .25in
Notes: Full Names: RXJ0228.2+2811; RXJ0751.3+5012; RXJ0819.6+6336; RXJ1205.1+3920; RXJ1320.1+3308; RXJ1532.9+3021; RXJ1715.3+5725; RXJ1740.5+3539 RXJ1750.2+3505\\

Total fluxes in parentheses are not significant.\\

Abell 1758b was eliminated due to confusion from Abell 1758a.  There were processing problems with the RXJ1532.9+3021 field, so it was not included in the sample.


\clearpage

\begin{figure}[h]
\begin{center}
\includegraphics[height=6cm]{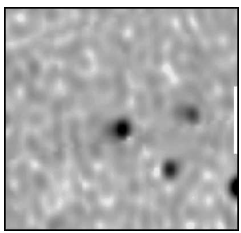}
\vskip 0.2in
\includegraphics[height=6cm]{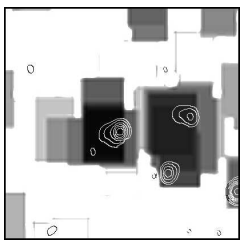}
\vskip 0.2in
\includegraphics[height=6cm]{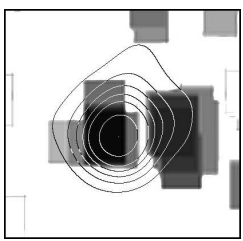}
\end{center}
\caption{Example of processing and detection of diffuse source in Abell 980.
Top: original WENSS image.  Middle: filtered WENSS image, showing diffuse emission in greyscale,  overlaid
by original image contours.   Bottom: filtered WENSS image overlaid by contours from ROSAT image smoothed to 240''.}
\label{A980}
\end{figure}

\begin{figure}[h]
\begin{center}
\includegraphics[height=9cm]{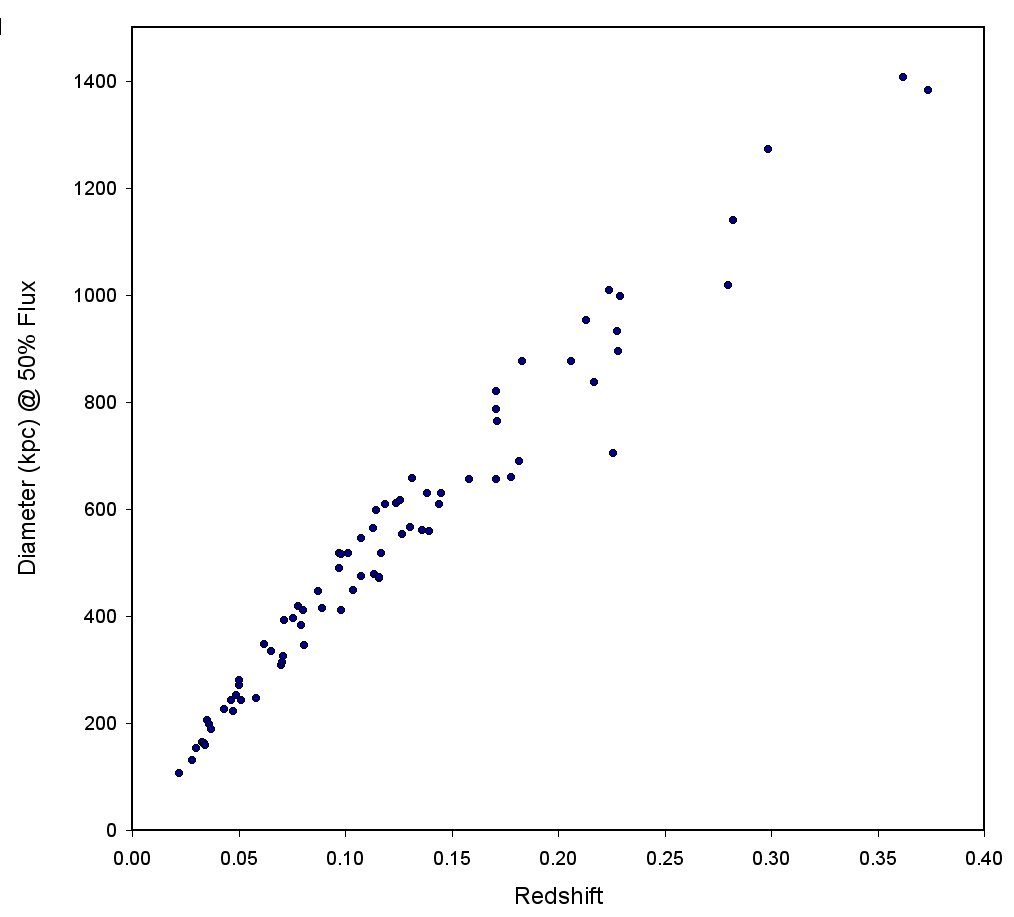}
\includegraphics[height=9cm]{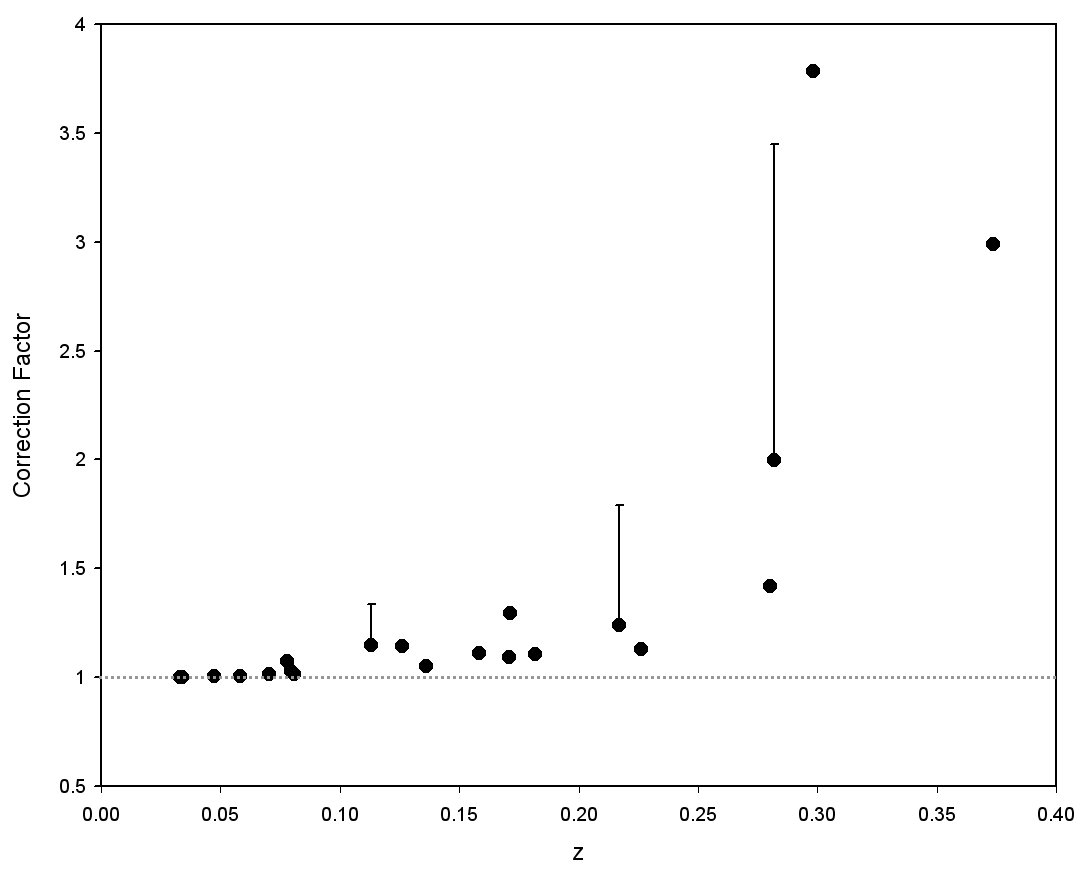}
\end{center}
\caption{Top: Diameter, as a function of redshift,  at which half of the diffuse emission would be included as part of the Mpc-scale flux in this survey. This shows both the magnitude of contamination from smaller scale structures and the loss of true Mpc-scale  diffuse emission. The variation at each redshift is due to the different declinations (beam sizes) of the clusters. Bottom: the correction factor for a 1 Mpc diameter halo as a function of redshift. Illustrative error bars indicate the correction needed if the halo had the same area, but was elliptical with a major/minor axis ratio of 3.}
\label{diam}
\end{figure}

\begin{figure}[h]
\begin{center}
\includegraphics[width=7cm]{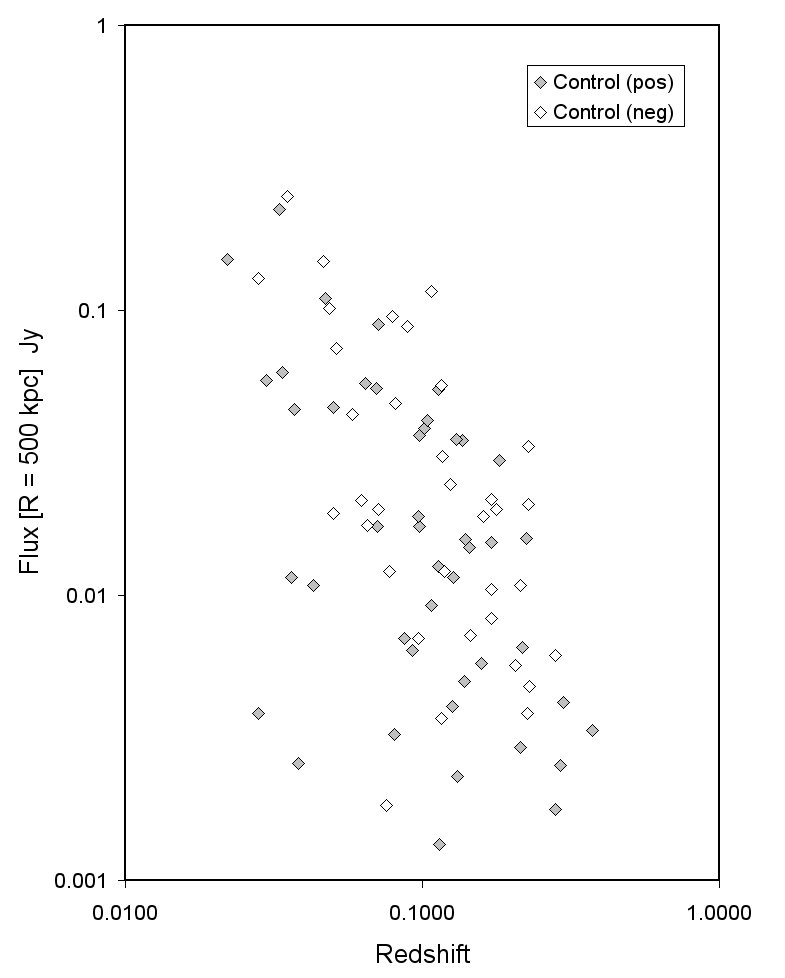}
\includegraphics[width=7cm]{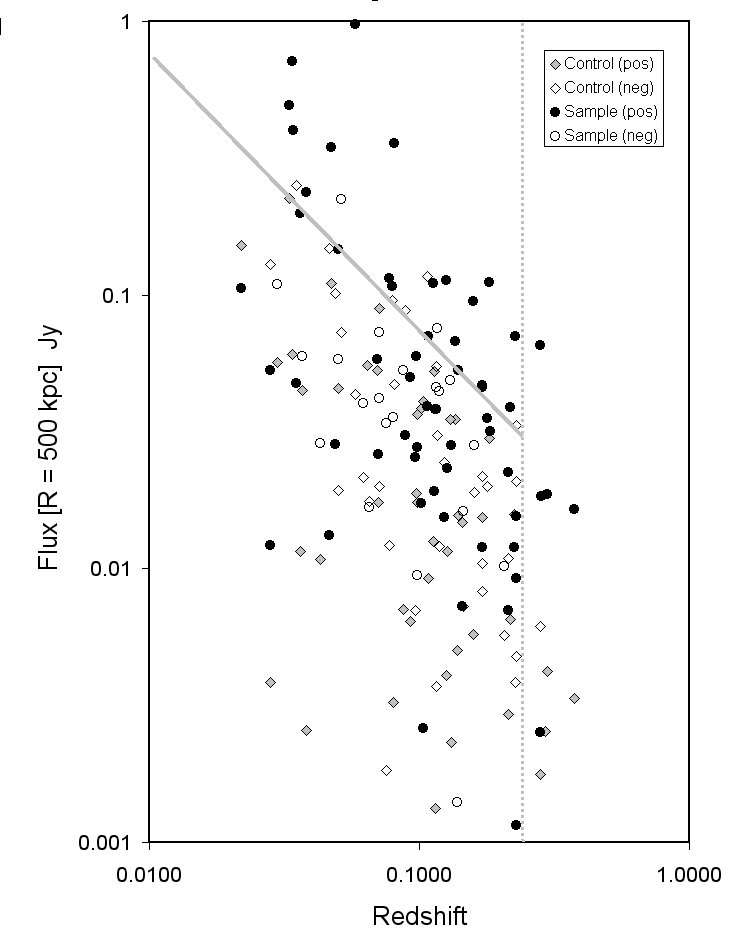}
\end{center}
\vskip -.25in
\caption{Observed flux within 500 kpc radius for control fields alone (left), and for control and sample fields (right), as a function of redshift.   The solid grey line  is the adopted 96\% confidence limit for the sample, with the sources at z$>$0.25 discussed in the text. Positive and negative measurements are plotted in absolute value.}
\label{control}
\end{figure}

\begin{figure}
\begin{center}
\includegraphics[width=12cm]{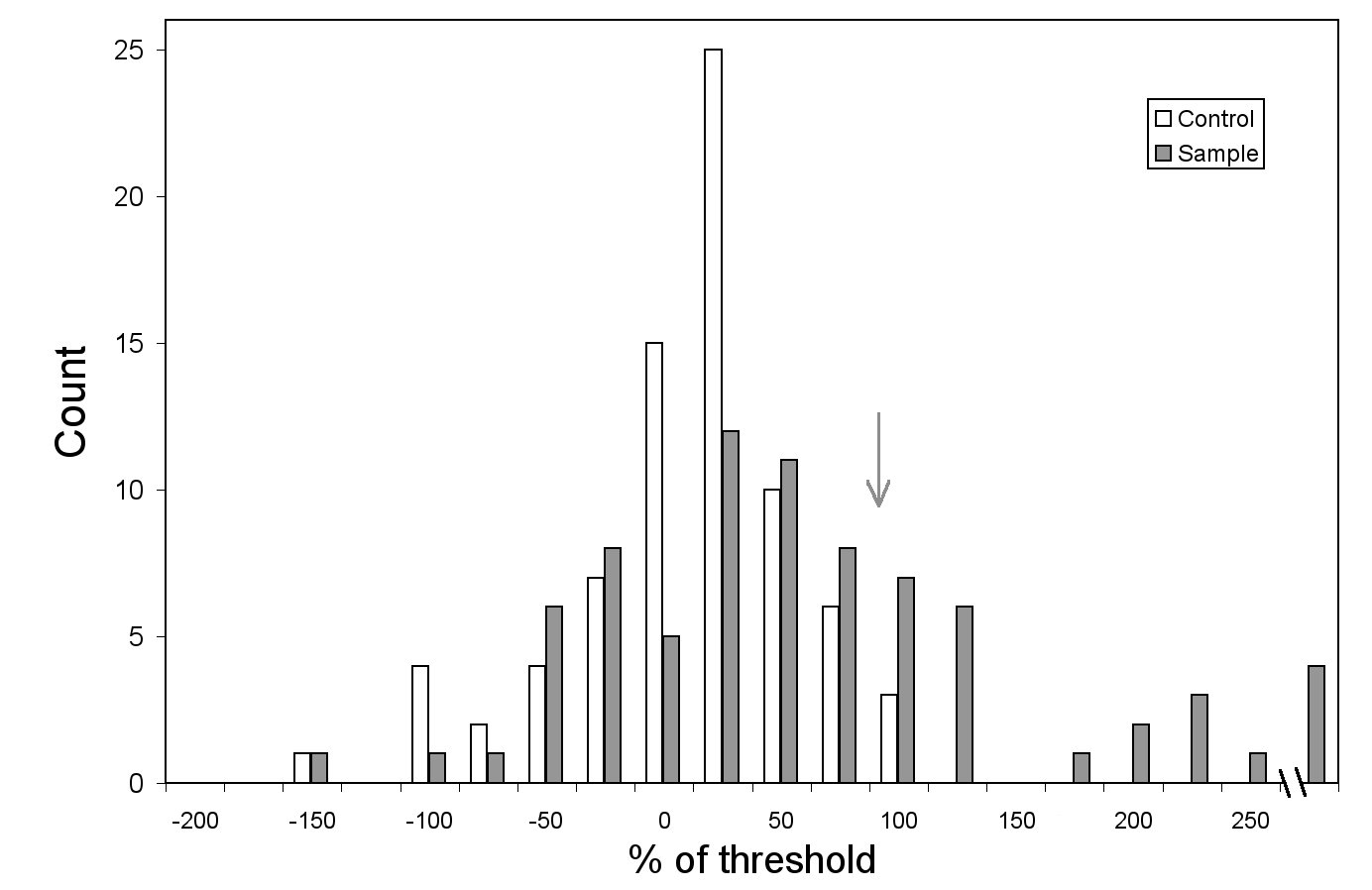}
\vskip 0.2in
\end{center}
\vskip -.25in
\caption{Count of the ratio (observed flux)/(threshold flux) for control and sample sources, in percent.  The arrow marks the adopted detection threshold.}
\label{distribute}
\end{figure}

\begin{figure}
\begin{center}
\includegraphics[width=6cm]{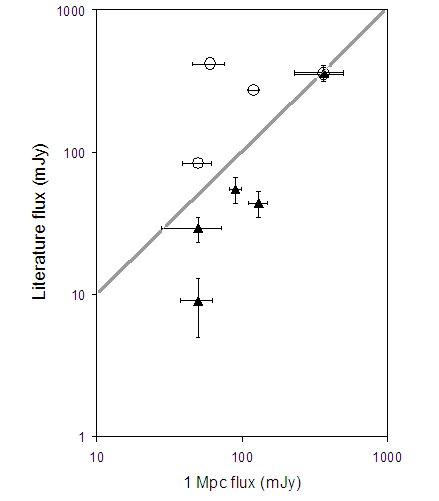}
\vskip 0.2in
\end{center}
\caption{Comparison of halo flux measurements from the current paper (defined as 1~Mpc) with those derived from the literature.  Solid triangles are values taken from \cite{kemp01}; open circles converted to 330~MHz and to fluxes from \cite{cass07}.  Error bars given in the latter paper are smaller than the sample size.}
\label{lit}
\end{figure}

\begin{figure}[h]
\begin{center}
\includegraphics[width=8.5cm]{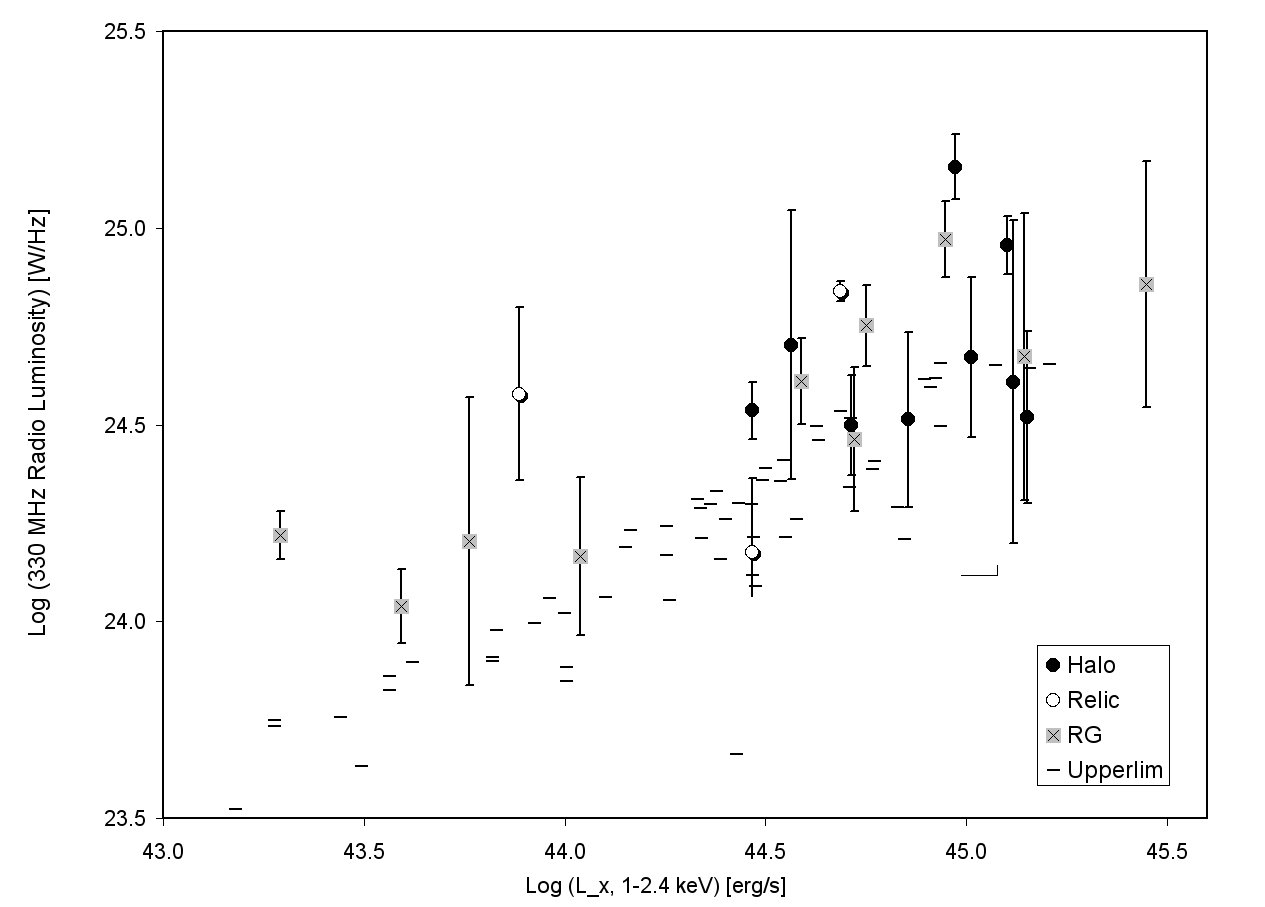}
\vskip 0.2in
\includegraphics[width=8.5cm]{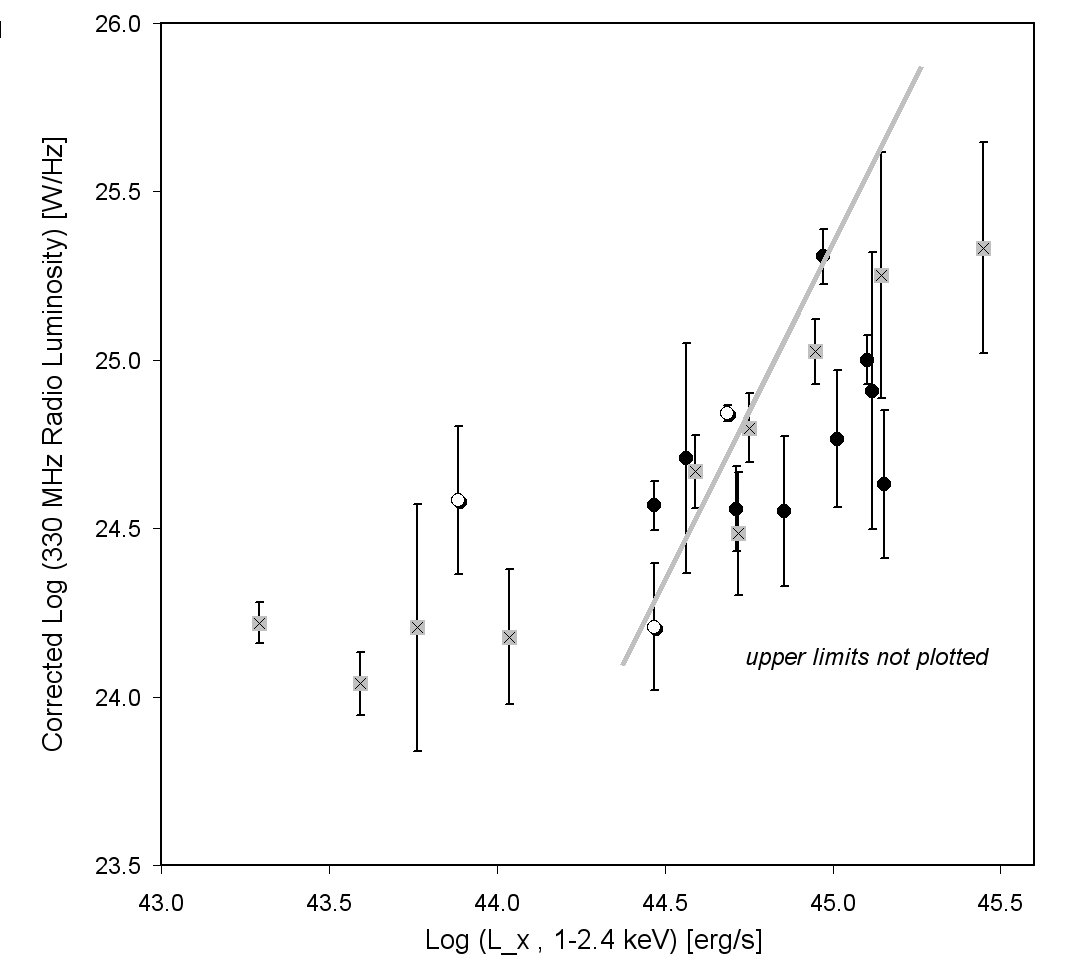}
\end{center}
\caption{Radio vs. X-ray luminosity for the sample. Filled circles: halos;  open circles: relics; grey boxes: radio galaxies; dashes: 96\% upper limits.  Top:  as observed. Bottom: Radio values corrected for the loss of diffuse flux in the filtering procedure.  No upper limits are shown. The solid grey line is the correlation shown by \cite{cass07}, scaled to 330 MHz with an assumed spectral index of -1. } 
\label{XvR}
\end{figure}

\begin{figure}[h]
\begin{center}
\includegraphics[height=9.5cm]{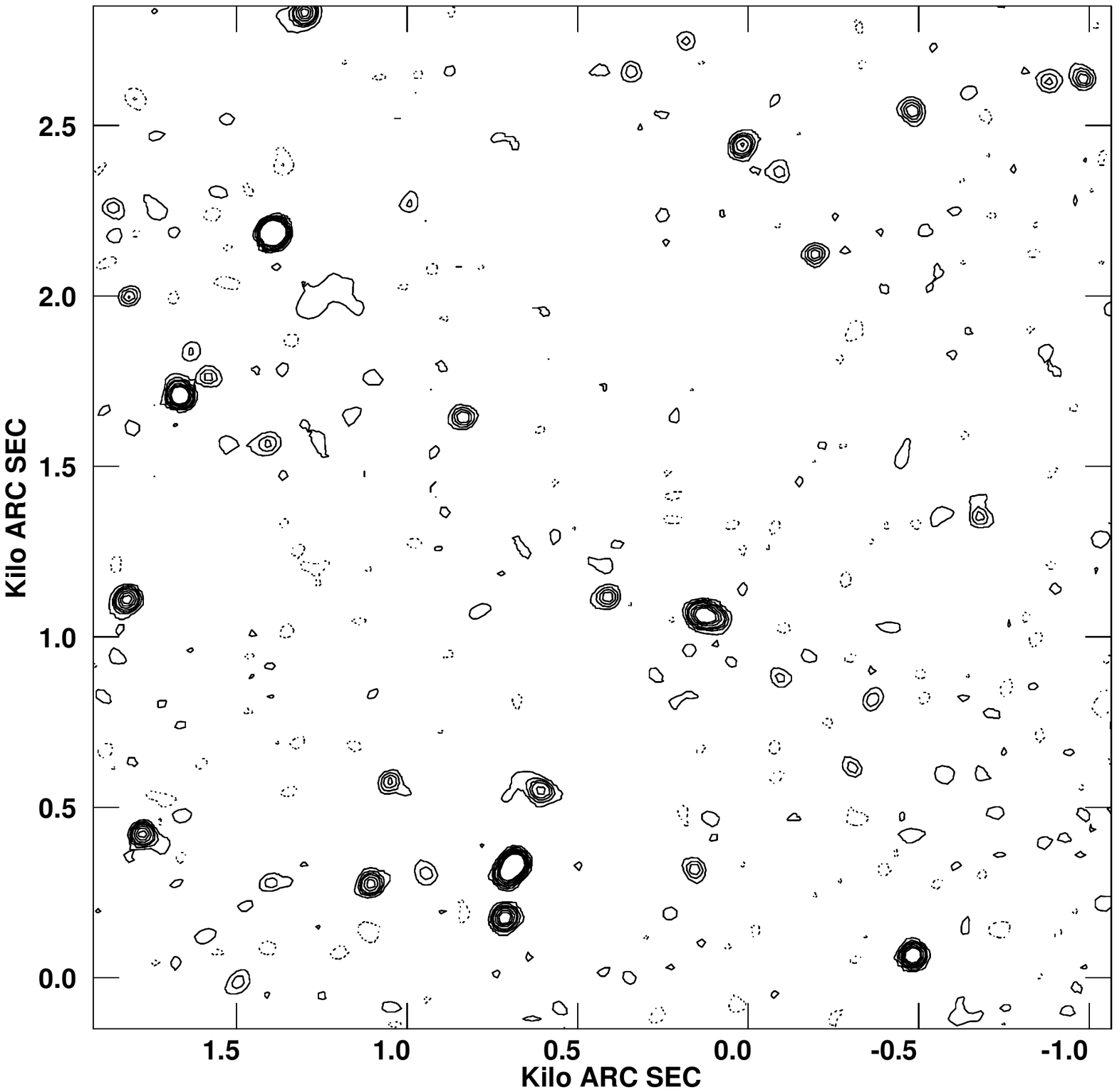}
\vskip 0.1in
\includegraphics[height=9.5cm]{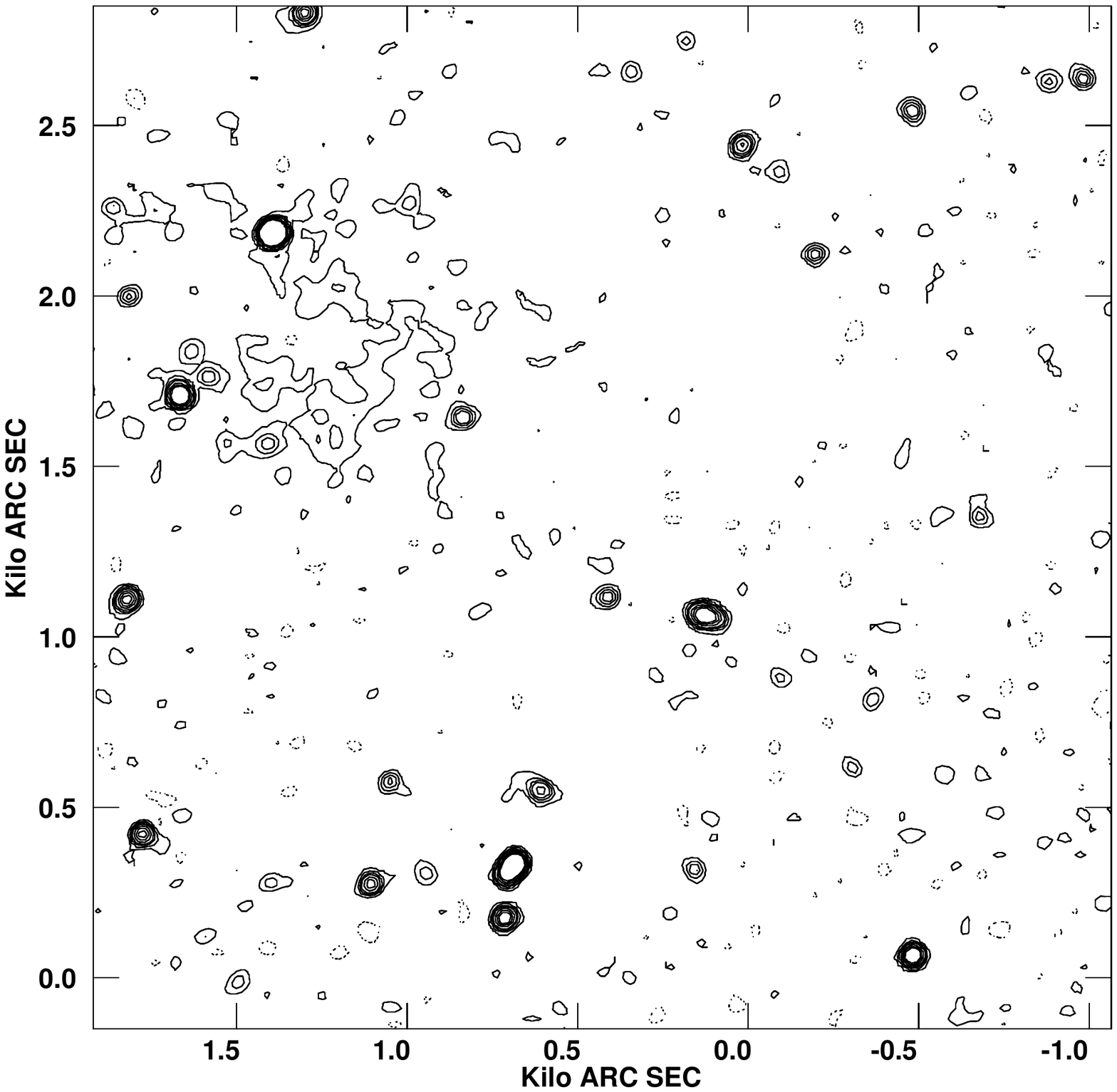}

\end{center}
\caption{Illustration of visibility of equal peak surface brightness objects as a function of angular size. Gaussian models of peak brightness 0.7 mJy/45'' beam have been added to an NVSS field at 80$^o$ galactic latitude. Top (bottom): FWHM = 300'' (900''). Contour levels are at 1 mJy/beam $\times$[-1,1,2,4,...]. Note that the NVSS is actually not sensitive to 900'' smooth structures; this figure is only for illustrative purposes.}
\label{contours}
\end{figure}

\begin{figure}[h]
\begin{center}
\includegraphics[width=12cm]{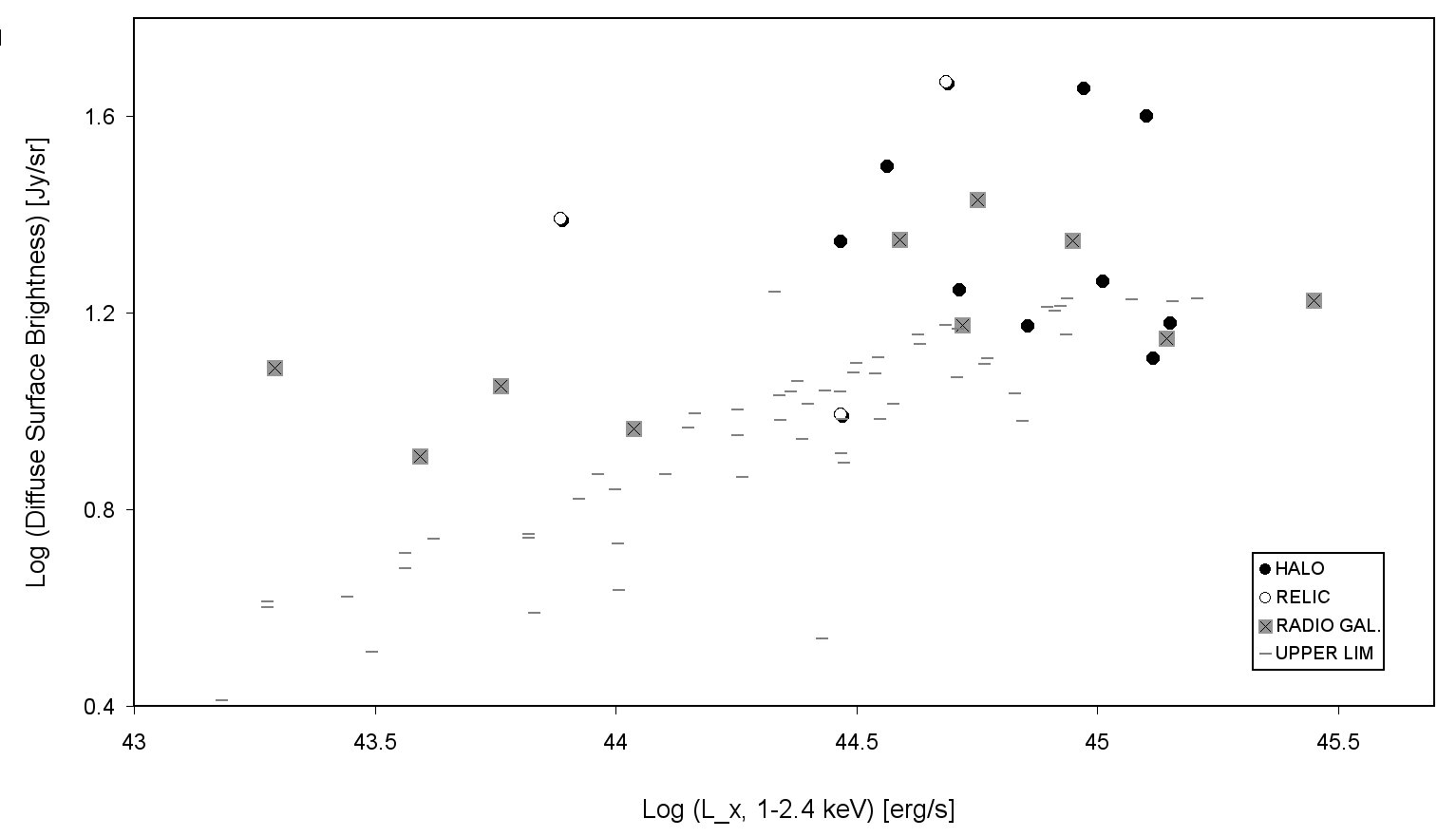}
\vskip 0.3in
\includegraphics[width=12cm]{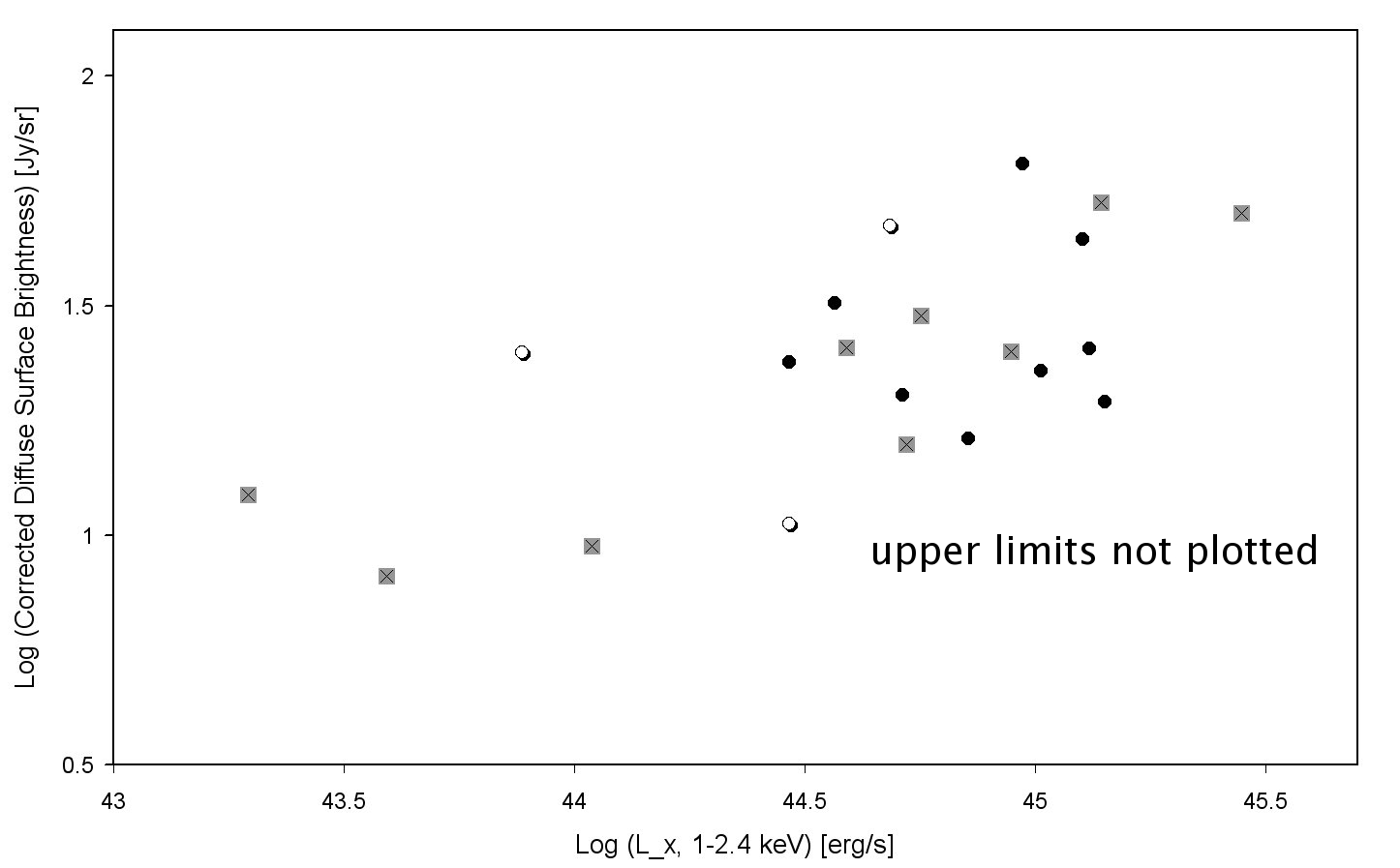}

\end{center}
\caption{Top: Observed radio surface brightness vs. X-ray luminosity for the current sample. Bottom: radio surface brightness for detections only,  corrected for the fractional loss of diffuse flux due to filtering.}
\label{sbright}
\end{figure}

\clearpage

 \section{Appendix - notes on individual clusters}

In this Appendix, we note some of the new information on individual clusters.  We do not provide a comprehensive review of the available data in the literature.

\subsection{Halo/Relic Systems}

\begin{figure}
\begin{center}
\includegraphics[width=13cm]{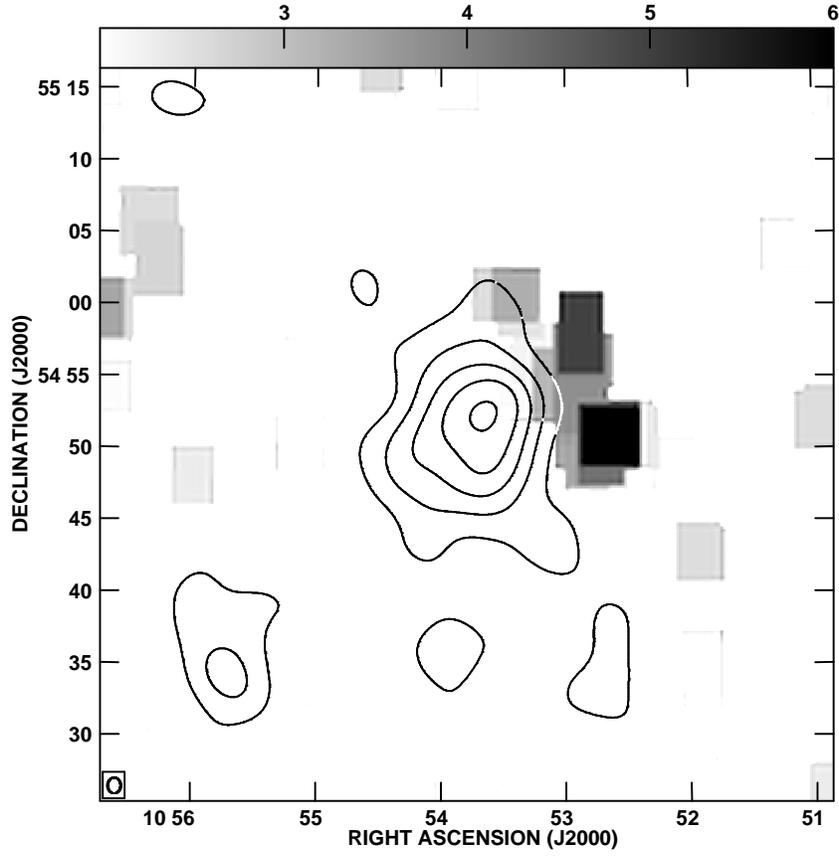}
\end{center}
\caption{RXJ 1053.7+5450, diffuse emission in greyscale, superposed by contours of ROSAT broadband counts, convolved to 240''. Contour levels are at 0.15*[5, 7, 9, 11, 13, 15] counts. }
\label{RXJ1053}
\end{figure}

{\it RXJ 1053.7+5450}~ ~ (Figure \ref{RXJ1053})\\
This new apparent relic is over 1 Mpc in extent, subtending over 90$^o$ of the cluster periphery. The southern bright patch in the relic is adjacent to a compact radio source, although there is no obvious association. This system deserves more detailed mapping. The continuing infall into this cluster was measured by \cite{rines06} using Sloan data.

\begin{figure}
\begin{center}
\includegraphics[width=8.5cm]{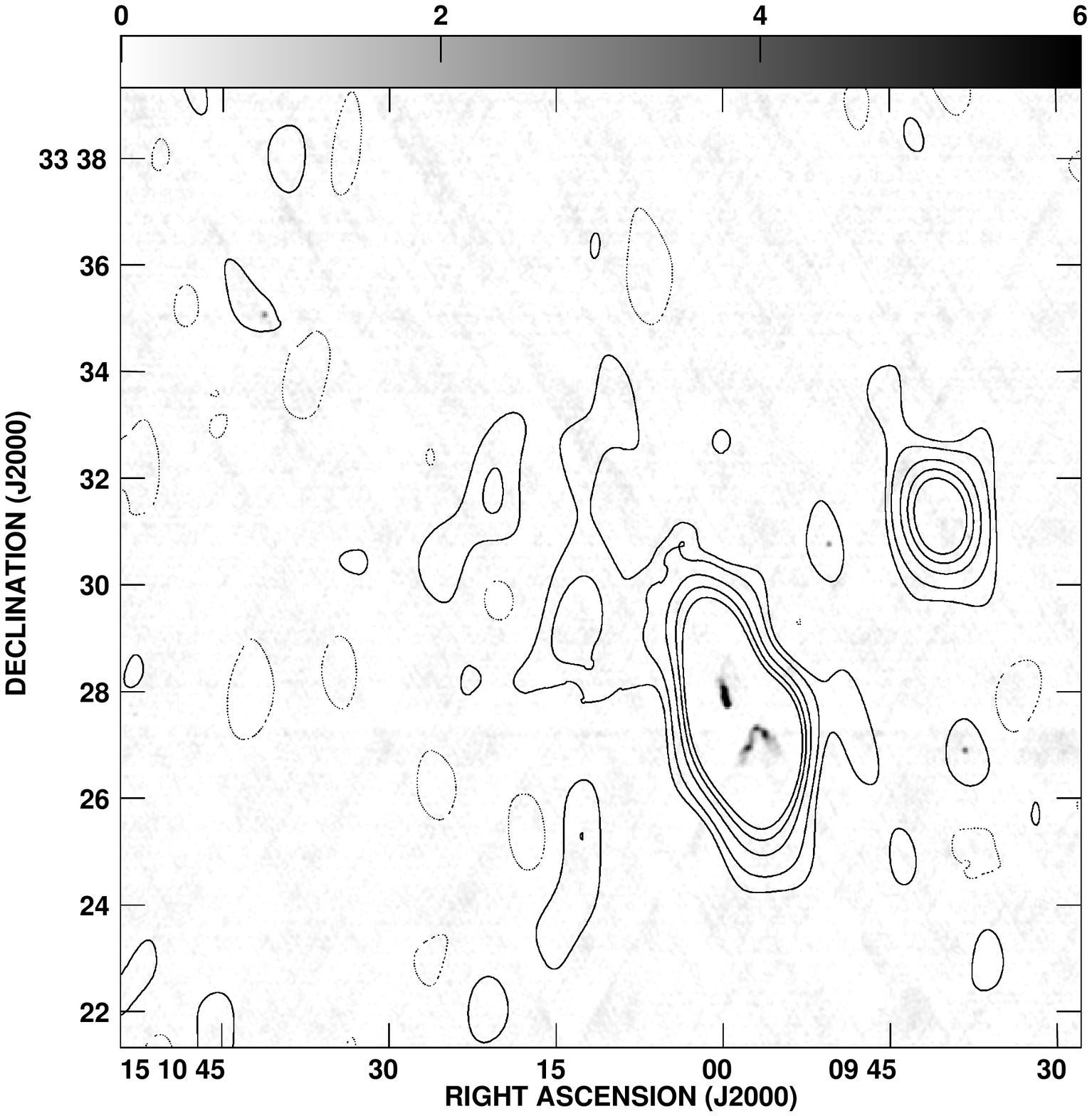}
\includegraphics[width=8.5cm]{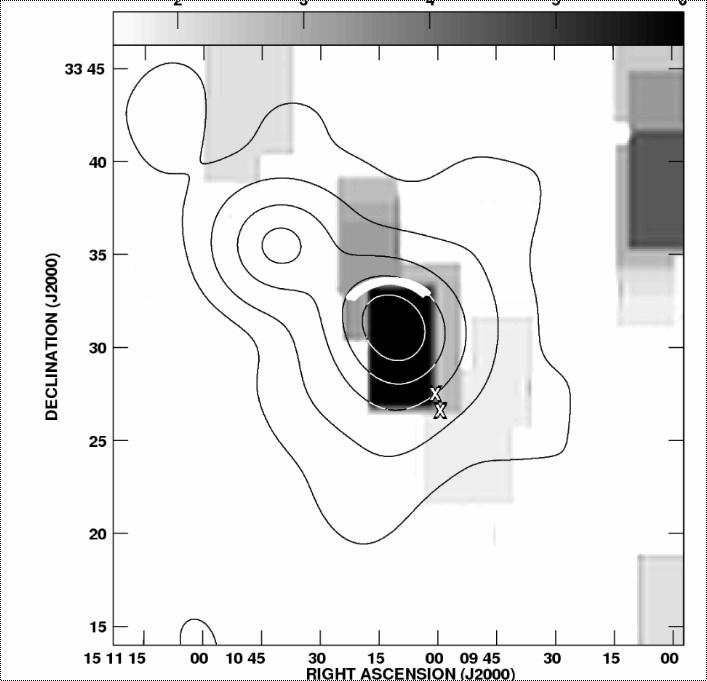}
\end{center}
\caption{Abell 2034.  Top: original WENSS contours at 7 mJy/beam * (-1,1,2,4,6,8).  Grey scale: FIRST image.  Bottom: Wider field, showing diffuse WENSS emission in greyscale, overlaid by ROSAT broadband contours, after convolution to 240'', at levels of 0.3 counts * (2,4,8,12,16,20). The position and approximate sizes of the radio galaxies are indicated by the white X's, and of the cold front by the white curve.}
\label{A2034}
\end{figure}

{\it Abell 2034}~ ~ (Figure \ref{A2034})\\
This diffuse feature was tentatively identified as a relic by \cite{kemp01}.  Using Chandra observations, \cite{kemp03} found multiple  signatures of an ongoing merger  including a cold front, and possible indications of Inverse Compton radiation from the radio source. Here we designate the diffuse source as a halo, because it is centered on the overall X-ray surface brightness, not at the cold front. We also show in Figure \ref{A2034} the overlay between the original WENSS image and the FIRST image at 5'' resolution.  Just SW of the X-ray peak there is a pair of tailed radio sources.  The southern one is a WAT, and the northern is a NAT that undergoes an extremely rapid expansion and drop in brightness as it extends toward the cluster center, where it can still be seen in WENSS, but not in FIRST.  It blends in to the more extended emission that we classify as a halo.The WENSS source to the NW of center is completely resolved out in FIRST. 

\begin{figure}
\begin{center}
\includegraphics[width=13cm]{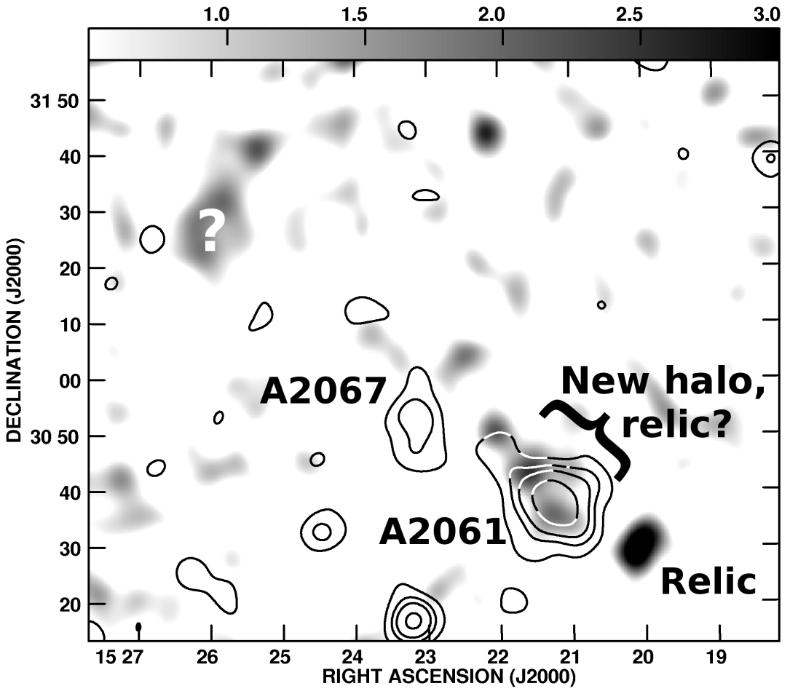}
\end{center}
\vskip -.25in
\caption{Abell 2061. Greyscale shows the diffuse radio emission, convolved to 300``, with ROSAT broadband contours at 80*(2, 3, 4, 6, 8) counts after convolution to 300''.  The new radio halo/relic system is visible along with the X-ray ``plume'' towards Abell 2067.  The bright diffuse radio feature marked with a {\emph ?} is not apparently associated with any other X-ray or radio structure.}
\label{A2061}
\end{figure}

{\it Abell 2061}~ ~ (Figure \ref{A2061})\\
The bright relic to the SW of the X-ray cluster emission was  found by \cite{kemp01}.  Here we show that there is additional diffuse radio emission coincident with the X-ray emission.  Both the X-ray and the diffuse radio emission extend to the NE, in the direction of A2067.  This system has been imaged using BeppoSAX MECS by \cite{marini}, who describe this extension as the ``plume'' and cite evidence for a shock within A2061 that may be caused by an infalling group.  \cite{rines06} note that with a separation of order Mpc in the sky, and a separation of only $\approx$ 600 km/s in redshift, Abell clusters 2061 and 2067 are probably bound. They are part of the Cor Bor supercluster \citep{small98}.

\begin{figure}
\begin{center}
\includegraphics[width=8cm]{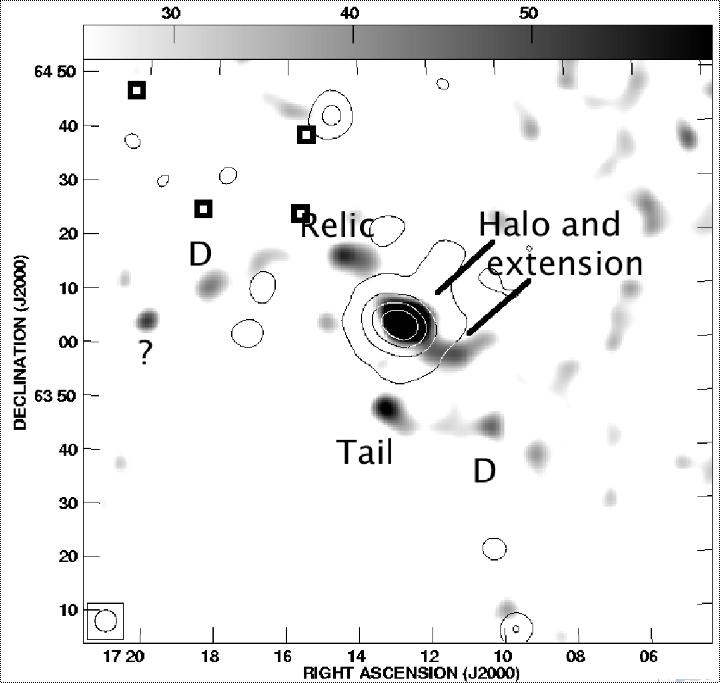}
\end{center}
\vskip -.25in
\caption{Abell 2255.  240'' convolution of greyscale diffuse radio emission and ROSAT broadband contours at intervals of 4 counts. In addition to the previously studied relic and halo, there is an extension of diffuse emission to the SW of the cluster, a radio tail to the south and two features marked ``D'' (see also \cite{pizzo}). Four SDSS clusters with similar redshifts \citep{miller} are marked as dark squares. The feature marked ? is diffuse emission around a bright compact radio source with no counterpart visible in SDSS.}
\label{A2255}
\end{figure}

{\it Abell 2255}~ ~ (Figure \ref{A2255})\\
This field shows the well-studied relic and halo emission and other low brightness but significant diffuse emission outside of the cluster core.  The very deep Westerbork image by \cite{pizzo} shows a large number of tailed and other radio galaxies threading the cluster.  In addition, there are patches of diffuse emission extending to the SW of the cluster (labeled as ``extension'' in Fig. \ref{A2255}), whose connection to individual radio galaxies is unclear. The features marked ``D'' are also detected by \cite{pizzo}. If all of these features are associated with Abell 2255, they span a diameter of $\approx$ 5 Mpc, and are thus approaching supercluster scales. Although not formally classified as a supercluster, there are four additional nearby clusters at very similar redshifts to the NE of Abell 2255 \citep{miller}.

\subsection{Radio galaxy + uncertain origins class}

Objects  were classified as ``radio galaxy'' because the diffuse emission  might arise from a low surface brightness extension of more compact radio emission.  In this subsection, we describe those sources worthy of more detailed study, because the association with radio galaxy emission is not clear, and there could be some actual halo or relic-type emission.  These are denoted by RG+? in Table 1.  Abell 980 has already been discussed in the main text and shown in Figure \ref{A980}.

\begin{figure}
\begin{center}
\includegraphics[width=8cm]{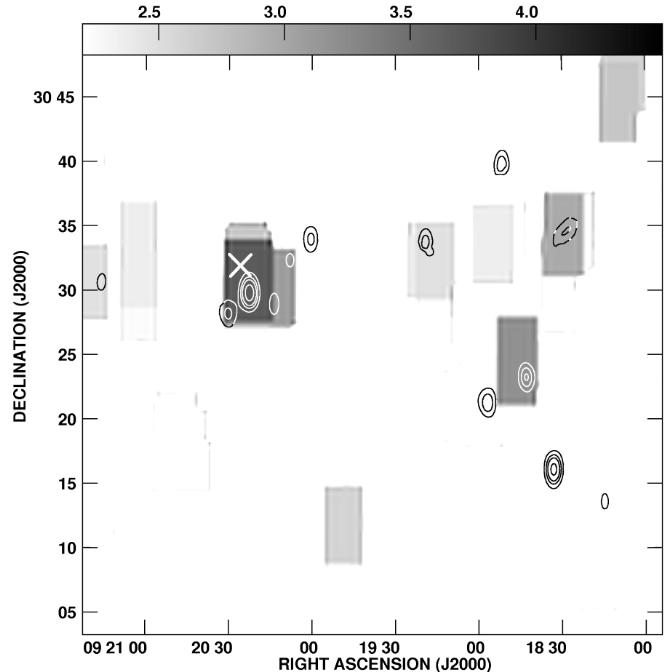}
\includegraphics[width=8cm]{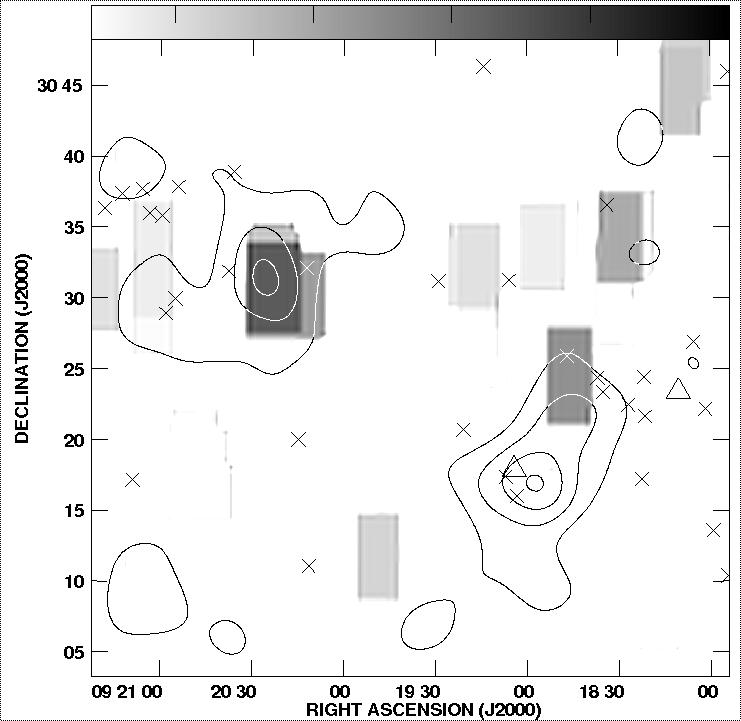}
\end{center}
\caption{Abell 781.  Left: Diffuse radio emission in greyscale overlaid by full resolution WSRT, with Abell 781 to the NE (X-ray peak position marked by white X).  Right:  Diffuse radio emission in greyscale overlaid by ROSAT contours, with Xs denoting galaxies from SDSS and 2MASS between z=0.12 and 0.14.  Triangles are the only clusters from \cite{gal03} in this redshift range (z$=$0.122 and 0.144). Abell 781 is at z$\sim$ 0.3, so the serendipitous SW structure is unrelated.}
\label{A781}
\end{figure}

{\it Abell 781}~ ~ (Figure \ref{A781})\\
The diffuse emission is centered somewhat to the north of the compact source, closer to the X-ray centroid, so there may be some halo contribution. Further to the east than studied here, \cite{vent08} have found a patch of diffuse radio emission at 610~MHz.   Serendipitously, we found an apparent extended X-ray source $\approx$20' to the SW of Abell 781. A bright region in the SW X-ray source is coincident with a QSO with a photometric redshift in SDSS of 0.145$\pm$0.001. Nearby, there are a number of galaxies at similar redshifts, and \cite{gal03} have identified two clusters in this area, NSC J091904+301755 ( photometric redshift 0.1439) and  NSC  J091810+302323 (photometric redshift 0.122). Scaling by the ROSAT counts in Abell 781, we calculate an approximate luminosity of the SW feature of 10$^{44.5}$ (10$^{44.1}$) erg/s with (without) the contribution from the QSO. Deep X-ray mapping is necessary to confirm the reality of diffuse emission outside of the QSO.


{\it Abell 1132}~ ~ (Figure 17)\\
The compact sources in this field contain a number of FRI radio galaxies \citep{odea85} as well as sources which are completely resolved away in FIRST and could use further mapping at intermediate resolutions.  \cite{gioferr00} also searched unsuccessfully for diffuse cluster emission using the VLA at 1.4 GHz.
Approximately 6' south of the cluster center, where the X-ray emission is falling off, is a source that appears to have a head-tail morphology at WENSS resolution. The peak of emission, also seen in NVSS, is coincident with an m$_R$=16.78 galaxy in SDSS (RA, Dec = 10 58 50.96, 56 43 08), with a redshift of  z=.139$\pm$.0002, the same as the Abell 1132 cluster.  The ``tail'' portion overlaps a group of galaxies at the same redshift,  $\approx$0.14, and has a length of 370 kpc.  After the AGN   turns off, the tail would appear to be a relic source, at a distance of 825 kpc from the cluster center. 





{\it A 1190}~ ~ (Figure \ref{abell1190})\\
The brightest radio source is a  tailed radio galaxy \citep{rud77}, and it is unclear whether there is additional diffuse emission \citep[see also][]{gioferr00}.   To the south, there is a 100'' extended source, that is fully resolved in  FIRST, and identified with a cluster galaxy. 

{\it RXJ1733.0+4345}~ ~ (Figure \ref{R1733})\\
This cluster (around IC1262) has been studied using BeppoSax by \cite{huds03} and \cite{huds+03} who suggest there may be diffuse non-thermal emission from a merger shock.  They describe the radio emission as a likely halo, but the combination of images shown here makes it likely that it is a 400 kpc WAT, with most of the emission resolved away at the 5'' resolution of FIRST.


\clearpage

\vskip -0.25in

\begin{figure}
\begin{center}
\includegraphics[height=6cm]{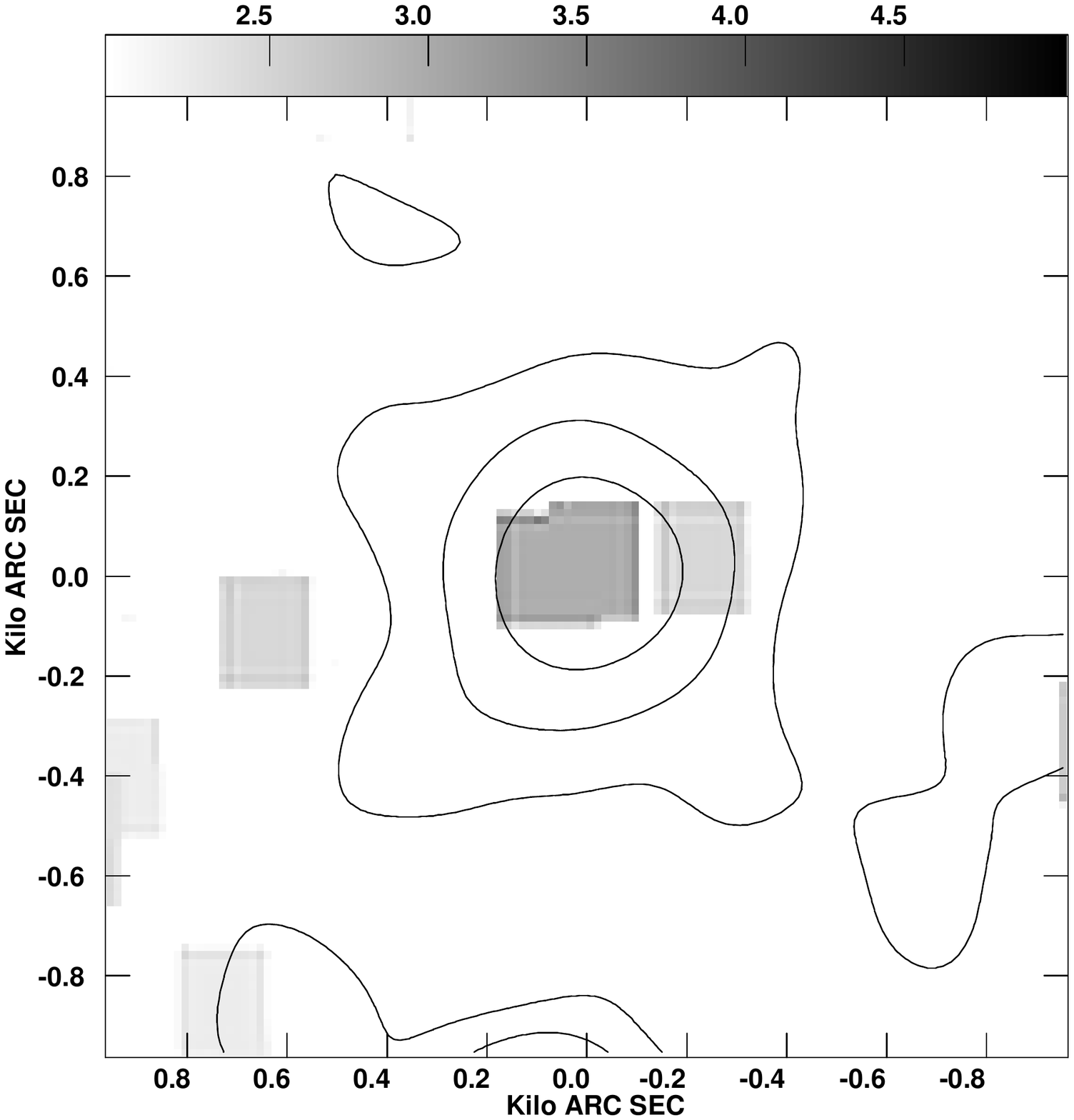}
\vskip 0.05in
\includegraphics[height=6cm]{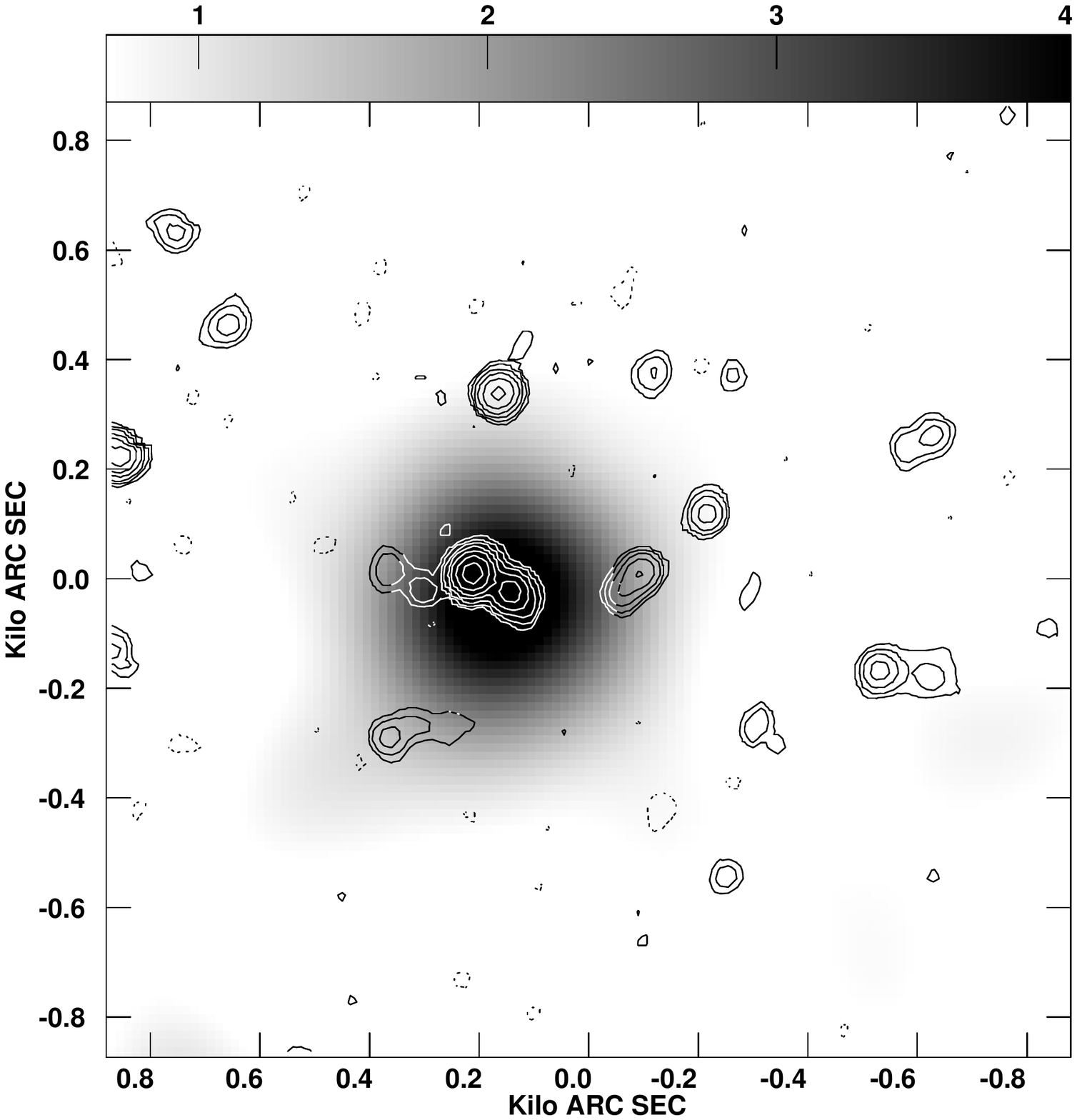}
\vskip 0.05in
\includegraphics[height=6cm]{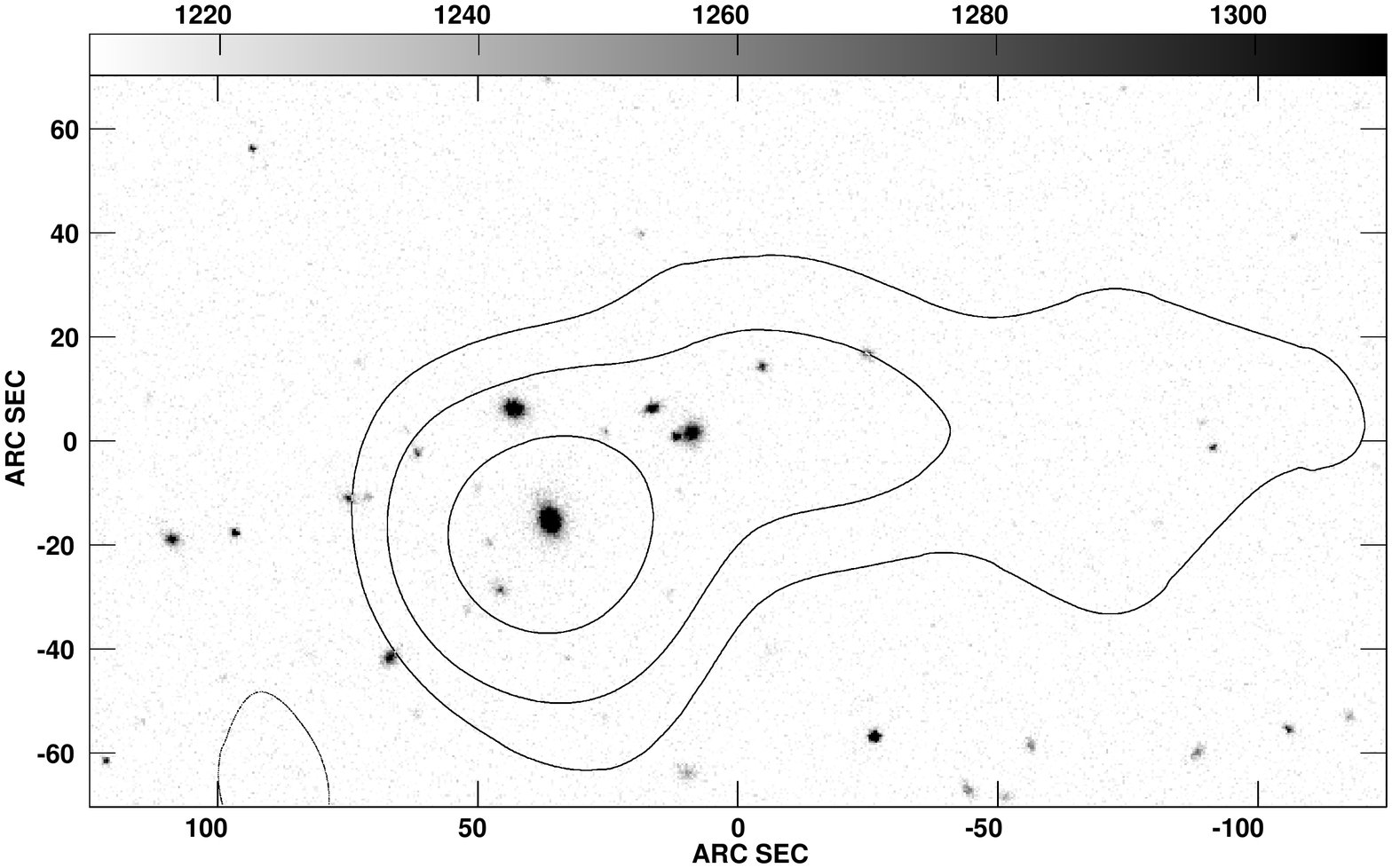}
\caption{Abell 1132. The top figure is a greyscale of diffuse radio emission overlaid with ROSAT contours.  The full resolution WENSS map is shown in the middle, overlaying ROSAT emission in greyscale.  South of the cluster is an apparent head-tail source, that is shown in more detail in the bottom image from the NVSS, overlaid on the optical field. The RA, Dec centers of the three maps, from top to bottom, respectively, are:
(10 58 26.014, 56 57 10.2), (10 58 06.839, 56 47 55.1), (10 58 46.628, 56 43 23.6).}
\end{center}
\label{A1132}
\end{figure}

\begin{figure}
\begin{center}
\includegraphics[height=13 cm]{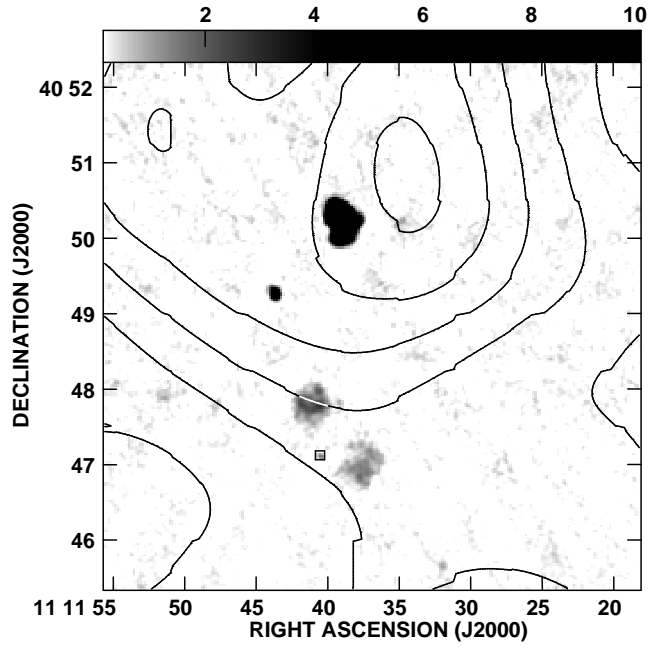}
\end{center}
\caption{FIRST image of Abell 1190 field, with the bright tailed radio galaxy near the peak of the ROSAT contours in the north.  The new low surface brightness radio source is visible as two diffuse lobes surrounding a weak central component to the south; its identification from SDSS, marked with a square,  is an m$_R$=15.38 galaxy, at a redshift of 0.074 $\pm$0.0002 at RA, Dec 11 11 40.5, 40 47 07.  The peak flux (black) is 10~mJy/5'' beam.    Contours are ROSAT broadband counts, after convolution to 135``, at levels of 0.25*(2,3,4,5,6,7).}
\label{abell1190}
\end{figure}

\begin{figure}
\begin{center}
\includegraphics[height=14cm]{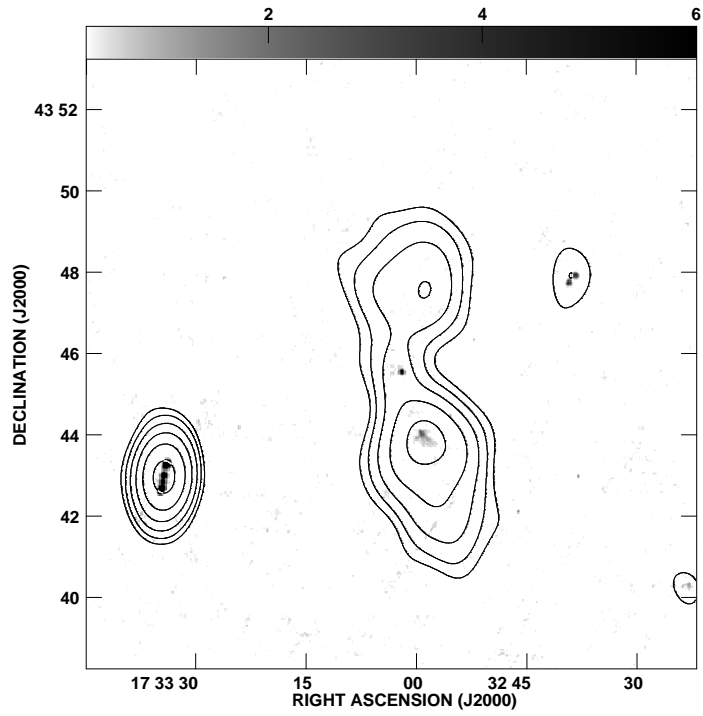}
\end{center}
\caption{IC1262 cluster. Greyscale of FIRST emission, overlaid with WENSS contours, at 0.02Jy/(78''$\times$54'' beam) * (-1,1,2,4,8,16,32), showing that the diffuse radio emission is likely a 400~kpc WAT.}
\label{R1733}
\end{figure}

\end{document}